\documentclass[review]{elsarticle}

\makeatletter
\def\ps@headings{%
\def\@oddhead{\mbox{}\scriptsize\rightmark \hfil \thepage}%
\def\@evenhead{\scriptsize\thepage \hfil \leftmark\mbox{}}%
\def\@oddfoot{}%
\def\@evenfoot{}}
\makeatother
\usepackage{amssymb}
\usepackage{graphicx}

\def\bq{\begin{equation}}
\def\eq{\end{equation}}
\def\bqn{\begin{eqnarray}}
\def\eqn{\end{eqnarray}}
\def\bqnn{\begin{eqnarray*}}
\def\eqnn{\end{eqnarray*}}

\def\bfone{{\bf 1}}
\def\defeq{\stackrel{\mathrm{def}}{=}}

\def\sensor_id{{\mathbb K}}
\def\measure{{\mathbb M}}
\def\est_set{{\mathbb S}}
\def\estcan_set{{\mathbb T}}

\newtheorem{remark}{\bf{Remark}}

\newcommand{\sizex}[1]{| #1 |_2}
\newcommand{\lengthx}[1]{| #1 |_1}

\begin{document}

\begin{frontmatter}
\title{Theoretical Framework for Estimating Target-Object Shape by Using Location-Unknown Mobile Distance Sensors}
\author{Hiroshi~Saito}
\author{Tatsuaki Kimura}

\address{NTT Network Technology Labs.\\3-9-11, Midori-cho, Musashino-shi, Tkyo 180-8585, Japan\\
tatsuaki.kimura.wa@hco.ntt.co.jp, kimura.tatsuaki@lab.ntt.co.jp}

\begin{abstract}
This paper proposes a theoretical framework for estimating a target-object shape, the location of which is not given, by using mobile distance sensors the locations of which are also unknown. Typically, mobile sensors are mounted on vehicles. Each sensor continuously measures the distance from it to the target object. The proposed framework does not require any positioning function, anchor-location information, or additional mechanisms to obtain side information such as angle of arrival of signal. Under the assumption of a convex polygon target object, each edge length and vertex angle and their combinations are estimated and finally the shape of the target object is estimated.

To the best of our knowledge, this is the first result in which a target-object shape was estimated by using the data of mobile distance sensors without using their locations.

\end{abstract}

\end{frontmatter}

\date{}

\section{Introduction}
Various distance sensors such as mm-wave sensors are implemented in cars to prevent traffic accidents and improve the comfort of driving. Because some of these sensors have ranges larger than 100 meters, they can gather environment information. This environment information is used by the car itself and can be useful even for other cars or people. If such information is used by other people for other applications, this is called vehicular-based participatory sensing or crowd sensing. 

Ideally, vehicular-based participatory sensing should be implemented without using location information in order to protect location privacy. (Although location privacy has been widely researched \cite{survey_privacy}, \cite{survey_privacy2}, it is not within the scope of this paper.) This study attempts to implement an application that estimates object shape without using location information. That is, without vehicles' location or moving direction information, we estimate the shape of a target object at an unknown location.

Such an estimation intuitively seems impossible due to there being too many unknown factors, and some theoretical results shown in the next section suggest it is impossible. However, by using mobile sensors that continuously measure the distance between individual sensors and the target object, this paper proposes a theoretical framework for successfully estimating the target-object shape. To the best of our knowledge, this is the first time a target-object shape has been estimated by using the data of mobile distance sensors without using their locations. This can be the first step to widely expanding the possibility of software sensors implemented by participatory sensing under complete location privacy. In addition, this paper also suggests that the secondary use of IoT (internet of things) \cite{IoT,IoTsurvey,commag,newIoTsurvey} information can be wider than expected.

The contributions of this paper are:
\begin{itemize}
\item This paper proposes a theoretical framework for estimating the shape of a convex polygon target object at an unknown location by using distance sensors the locations of which are not given. Each sensor moves on an unknown line at a known speed and continuously measures the distance from it to the target object. The estimation framework does not require any positioning function, anchor-location information, or additional mechanisms to obtain side information such as angle of arrival of signal. 

\item The estimation problem includes a unique aspect: the sensing information includes unknown factors. That is, neither the sensor's location nor its moving direction is given. The proposed framework estimates each part of the target-object shape and the combinations of each part. This is a new type of estimation algorithm.
\end{itemize}

\section{Related work}
The fundamental questions related to the research topic of this paper is whether we can estimate the shape of a target object by using many simple sensors such as distance sensors or binary sensors without a positioning function or location information and how we estimate it if possible. Our prior studies suggested that we can estimate only a small number of parameters such as the size and perimeter length of a target object by using randomly deployed sensors such as binary sensors and distance sensors and cannot estimate other parameters \cite{infocom,ieice-invite,arXiv}. Thus, these studies introduced composite sensors that are composed of several simple sensors and are randomly deployed. By using them, additional parameters were able to be estimated \cite{signalProcess,mobileComp}. The studies used the sensing results at a certain sensing epoch and estimated parameters using them. Even when the studies used the sensing results at multiple sensing epochs, they did not take into account sensing epoch information. Only one study \cite{time-variant} among these studies took account of sensing epochs and the temporary behavior of sensing results, but it focused on estimating the size and perimeter length of the target object. Recently, we have developed a framework for estimating the shape of a target object moving on a unknown trajectory at a unknown speed by using distance sensors at unknown locations \cite{fixed_sensor}.
The estimation method structure in which parts of the target object and their connectivities are estimated is similar, but there are major differences between this paper and that paper.
(i) The sensing area model in that paper is a special case of this paper.
(ii) The estimation in that paper needs to estimate the target object's moving speed.

As far as we know, no studies other than those mentioned above have directly tackled these questions. However, there has been a considerable amount of studies on developing an estimation method that uses location-unknown sensors. These studies took a different approach. Most first estimated the sensor locations \cite{locating_nodes} because it is believed that ^^ ^^ the information gathered by such sensor nodes will generally be useless without determining the locations of these nodes" \cite{flip_amb} or ^^ ^^ the measurement data are meaningless without knowing the location from where the data are obtained" \cite{local_4}. Once sensors' locations are estimated, shape estimation is no longer difficult. However, an approach of estimating the sensor locations often requires additional mechanisms or side information, such as locations of anchor sensors, angle-of-arrival measurements, training data and period, and distance-related measurements \cite{locating_nodes,local_4,local_2,local_3}. Concrete examples are intersensor distance information \cite{flip_amb}, location-known anchor sensors \cite{tsp2002}, a set of signals between sensors \cite{acm_sensor}, and the system dynamic model and location ambiguity of a small range \cite{bernoulli}.

In addition, there has been a research into capturing the shape of a target object by using cameras that cannot cover the whole shape of the target object \cite{camera}.

\section{Model}
\begin{figure}[tb] 
\begin{center} 
\includegraphics[width=11cm,clip]{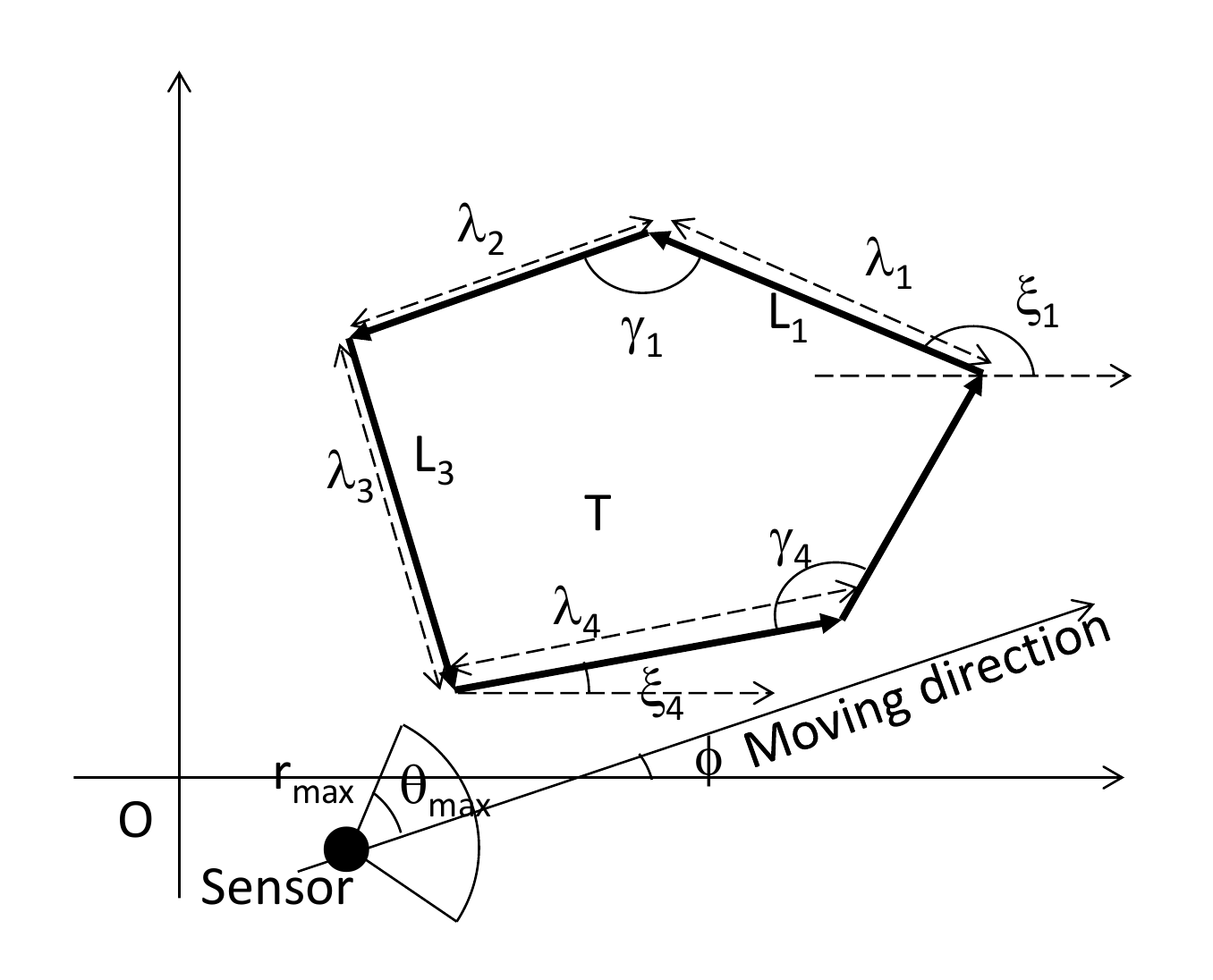} 
\caption{Illustration of target object model} 
\label{model} 
\end{center} 
\end{figure}

A fixed target object $T$ in a bounded convex set $\Omega\subset\mathbb{R}^2$ and is a convex polygon.
Its boundary $\partial T$ is closed and simple (no holes or double points) and consists of directional edges $\{L_j\}_j$ where $j =1,2,\cdots,n_e$ (Fig. \ref{model}).
Here, $n_e$ is the number of edges.
Let $\lambda_{j}$ be the length of $L_{j}$, and let $\xi_{j}$ be the angle formed by $L_{j}$ and the reference direction where $0\leq \xi_j<2\pi$.
Note that the inner angle formed by $L_{j}$ and $L_{j+1}$ is $\gamma_j=\pi-\xi_{j+1}+\xi_j$.
Here, $\{L_{j}\}_j$ are counted counterclockwise along $\partial T$, and the head of $L_{j}$ is the tail of $L_{j+1}$.
We do not know any of $\{\lambda_{j}, \xi_{j}, \gamma_j\}_{j}$.
That is, we do not know the target-object shape or size.

A vehicle is running at a speed $v$ on a randomly placed straight line the direction of which is $\phi$ from the reference direction where $\phi$ is an independent random variable uniformly distributed in $[0,2\pi)$.
That is, the vehicle's location (that is, the sensor's location) $(x_s(t),y_s(t))$ is given by $(vt\cos\phi+x_s(0),vt\sin\phi+y_s(0))$.
For simplicity, assume that $v$ is time-invariant.
However, the extension to a time-variant $v$ is straightforward.
The vehicle is equipped with a directional distance sensor and a speed meter measuring $v$.
(That is, $v$ is known, but $\phi$ is unknown.)

The distance sensor measures the distance from the sensor to the nearest point of $T$ within the sensing area.
(In practice, $(x_s(t),y_s(t))$ may not be able to be in $T$, but the vehicle is assumed to run on a straight line passing through $T$ for simplicity.)
Its sensing area is a sector shape of radius $r_{max}$ and direction range $[-\theta_{max},\theta_{max}]$ from its moving direction where $0<\theta_{max}\leq \pi/2$.
That is, the sensing area is $(x_s(t)+u\cos(\phi+\theta),y_s(t)+u\sin(\phi+\theta))$ for $0\leq \forall u\leq r_{max},-\theta_{max}\leq\forall\theta\leq\theta_{max}$.
The sensor continuously measures the distance $r(t)$ at $t$ from the sensor to the target object and sends the sensing result with $v$  to a server collecting sensing results from individual sensors.
Thus, $r(t)$ is given as follows: 
$
r(t)=\cases{\tilde r(t), &if $\tilde r(t)\leq r_{max}$,\cr
\emptyset,&if $\tilde r(t)>r_{max}$.}
$
Here, $\tilde r(t)\defeq\min_{(x_s(t)+u\cos(\phi+\theta),y_s(t)+u\sin(\phi+\theta))\in T, -\theta_{max}\leq\theta\leq\theta_{max}} u$.
In particular, $r(t)=0$ if $(x_s(t),y_s(t))\in T$.
Define the detecting direction $\theta^*$ and the detected point as follows.
$\theta^*$ is $\theta\in [-\theta_{max},\theta_{max}]$ minimizing $\{u|(x_s(t)+u\cos(\phi+\theta),y_s(t)+u\sin(\phi+\theta))\in T\}$ and the detected point is $(x_s(t)+r(t)\cos(\phi+\theta^*),y_s(t)+r(t)\sin(\phi+\theta^*))$ on $\partial T$ for $r(t)>0$.
The sensor continuously sends a report of $r(t)$ and $v$ to an estimation server.
(If $r(t)=\emptyset$, NO DETECTION is reported.)
That is, we can use $r(t)$ and $v$ of each sensor.
Neither the vehicle's location $(x_s(t),y_s(t))$ nor moving direction $\phi$ is given to protect location privacy.

There are $n_s$ vehicles monitoring $\Omega$.
$\phi,v,r(t)$ of the $i$-th vehicle or its sensor are described as $\phi_i,v_i,r_i(t)$.

Table \ref{p_list} lists the variables and parameters used in the remainder of this paper for the reader's convenience.
\begin{table}
\caption{List of variables and parameters}
\begin{center}\label{p_list}
\begin{tabular}{ll}
\hline
$T$&target object\\
$L_{j}$&$j$-th directional line segment of $\partial T$\\
$\lambda_{j}$&length of $L_{j}$\\
$\xi_{j}$&angle formed by $L_j$ and reference direction\\
$\gamma_j$&inner angle formed by $L_j$ and $L_{j+1}$\\
$n_e$&number of edges in $\partial T$\\
$n_s$& number of sensors\\
$r_{max}$&maximum sensing range\\
$\phi$&angle of vehicle's moving direction\\
$v$&moving speed of vehicle\\
$\theta_{max}$&sensing direction range from vehicle's moving direction\\
$\theta^*$&detecting direction\\
$r(t)$&measured distance to $T$ at $t$\\
$\zeta$&$\xi+\pi/2-\phi$\\
$\zeta_p$&$\xi+\pi/2+\theta_{max}$\\
$\zeta_m$&$\xi+\pi/2-\theta_{max}$\\
$p_d(L)$&period of $r(t)$ detecting a whole edge $L$\\
$l_d(L)$&length in time of $p_d(L)$\\
$s_d(L)$&slope of $r(t)$ during $p_d(L)$\\
$n_d(\lambda)$&number of sensors detecting the whole edge of length $\lambda$\\
$n_d(\gamma)$&number of sensors detecting a vertex of angle $\gamma$\\
$\estcan_set(x)$&set of candidate estimates derived from $x$\\
$k_{sub}$& ratio of total number of candidate estimates to number of sub-intervals\\
$c(\widehat{x})$&number of occurrences of estimates ($x=\lambda$ or $\gamma$)\\
$\est_set(x)$&set of estimates of edge length ($x=\lambda$) or angle ($x=\gamma$)\\
$n_\lambda$&number of whole edge detection samples\\
$n_\gamma$&number of vertex detection samples\\
$\widehat{N_\lambda}$&estimated number of edges of length $\lambda$\\
$\widehat{N_\gamma}$&estimated number  of vertexes of angle $\gamma$\\
\hline
\end{tabular}
\end{center}
\end{table}

In the remainder of this paper, we use the following notations. For a set $X\subset \mathbb{R}^2$, $\partial X$ denotes its boundary, $\lengthx{X}$ denotes its perimeter length, and $\sizex{X}$ denotes its area size. $\sharp(S)$ is the number of elements in a discrete set $S$, $\bfone(z)\defeq\cases{1, &if $z$ is true,\cr 0, &otherwise,}$, $\bfone_\emptyset(z)\defeq\cases{1, &if $z$ is true,\cr \emptyset, &otherwise,}$,$[z]^+\defeq z\bfone(z>0)$, and $\widehat{z}$ is an estimator of $z$. In addition, $\arcsin(t)$, $\arccos(t)$, and $\arctan(t)$ take values in $[-\pi/2,\pi/2)$, $[0,\pi)$, and $[-\pi/2,\pi/2]$, respectively.

\section{Basic properties}\label{basic_pro}
This section discusses basic properties of $r(t)$.

A sensor detecting $L_j$ with a fixed $\theta^*\in[-\theta_{max},\theta_{max}]$ needs to satisfy
\bq
\xi_j-\theta^*\leq\phi\leq\xi_j-\theta^*+\pi.\label{fix-theta-xi}
\eq

\begin{figure}[tbh] 
\begin{center} 
\includegraphics[width=8cm,clip]{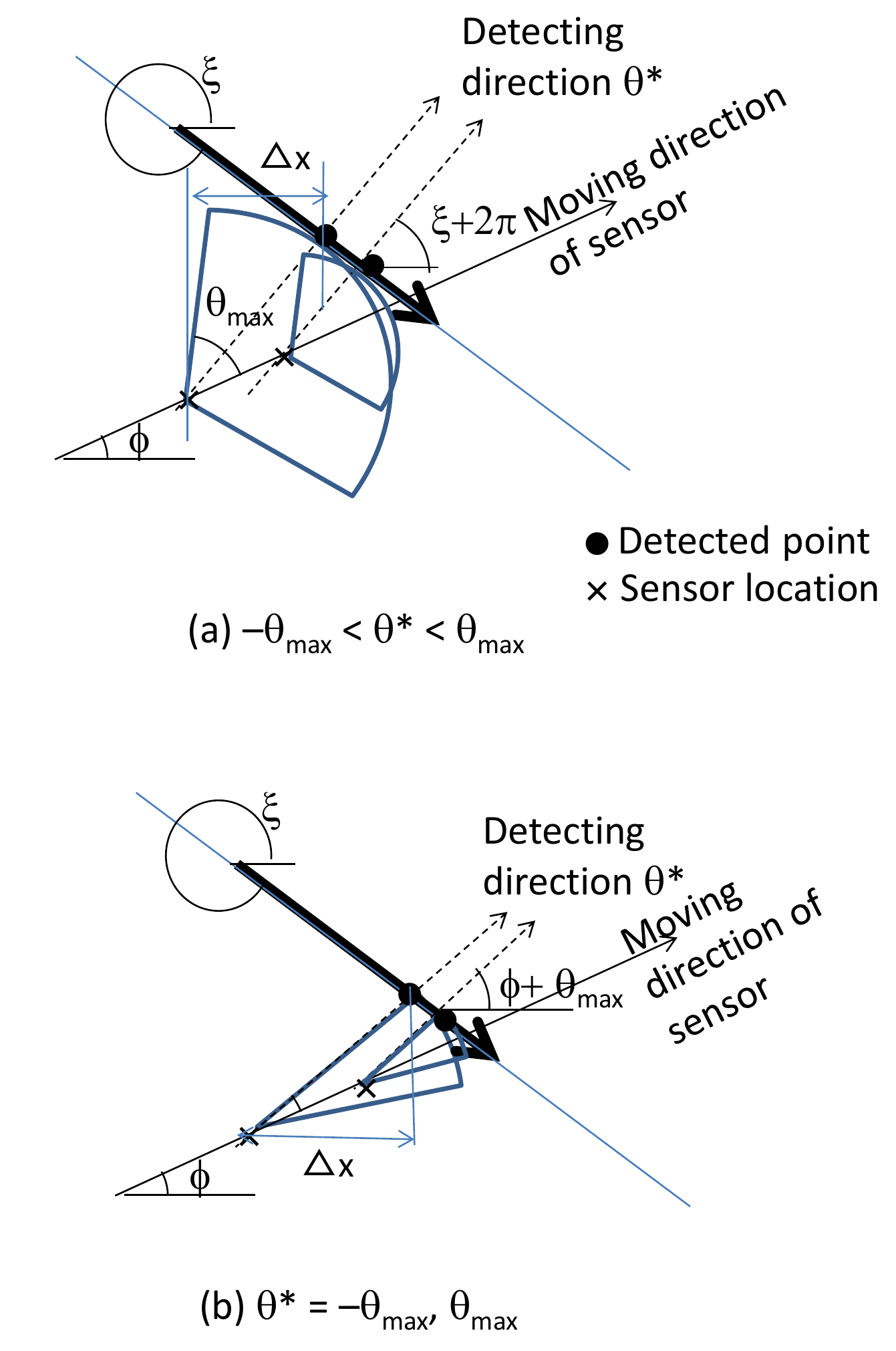} 
\caption{Illustration of detection} 
\label{basic} 
\end{center} 
\end{figure}

Note that the detecting direction $\theta^*$ is $\zeta\defeq\xi+\pi/2-\phi$ or $\pm\theta_{max}$ for an edge of direction $\xi$ when the detected point is not at an end of the edge (Fig. \ref{basic}).
When 
\bq
-\theta_{max}\leq \zeta\leq\theta_{max},\label{vert-d}
\eq
$\theta^*=\zeta$.
When 
\bqn
&&-\theta_{max}-\pi/2<\zeta<-\theta_{max} \cr 
&&(\theta_{max}<\zeta<\theta_{max}+\pi/2),\label{max-d}
\eqn
$\theta^*=-\theta_{max}$ ($\theta^*=\theta_{max}$).
Equivalently, 
\bq
\theta^*=\cases{
\theta_{max}, &for $\zeta_m-\pi/2<\phi<\zeta_m,$\cr
\zeta, &for $\zeta_m\leq\phi\leq \zeta_p,$\cr
-\theta_{max}, &for $\zeta_p<\phi<\zeta_p+\pi/2$.
}\label{detecting_direction}
\eq

Eq. (\ref{vert-d}) means that the sensor can detect the distance to the edge at the vertical direction of the edge.
Because the distance sensor normally detects the minimum distance to an object, this is a normal case.
Eq. (\ref{max-d}) means that the sensor cannot detect the distance of the edge at the vertical direction.
In this case, the detecting direction becomes $\pm\theta_{max}$, which is the closest direction to the vertical direction of the edge within the sensing direction range.

Consider a line on which an edge of direction $\xi$ exists. Without loss of generality, we can assume that this line passes through the origin. Then, this line can be expressed as 
\bq
y=(\tan\xi)x\label{line_xi}
\eq
on the $(x,y)$-coordinate.

\subsection{Relationship between $s_d, l_d$ and parameters of $T$}
\subsubsection{For $\theta^*=\zeta$}
When the detecting direction is $\theta^*=\zeta$, the line the direction of which is the same as the detecting direction and that passes through the sensor's location is 
\bq
y=(\tan(\xi+\pi/2))(x-tv\cos\phi-x_s(0))+tv\sin\phi+y_s(0).
\eq
The intersection $(x^*,y^*)$ of this line and the line defined by Eq. (\ref{line_xi}) (the edge of direction $\xi$ is on) is 
\bqn
x^*(t)&=&(tv\cos\phi+x_s(0))\cos^2\xi\cr
&&+(tv\sin\phi+y_s(0))\sin\xi\cos\xi,\\
y^*(t)&=&(tv\sin\phi+y_s(0))\sin^2\xi\cr
&&+(tv\cos\phi+x_s(0))\sin\xi\cos\xi.
\eqn
Thus, the relative location $(\triangle x(t), \triangle y(t))$ of this intersection from the sensor's location is $(x^*(t)-tv\cos\phi-x_s(0), y^*(t)-tv\sin\phi-y_s(0))$.
Because $r(t)=|\triangle x(t)/\cos(\xi+\pi/2)|$ when the intersection is on the edge (that is, the intersection becomes a detected point), $r(t)$ is a linear function of $t$ when the sign of $\frac{\triangle x(t)}{\cos(\xi+\pi/2)}$ is fixed and its slope $s_d$ (the amount of increase/decrease of $r(t)$ per a unit of time) is 
\bq
s_d= \pm v\sin(\xi-\phi).
\eq
Because $\xi-\phi$ must satisfy Eq. (\ref{vert-d}),
\bqn
&&\xi-\phi\cr
&=&\cases{\pm\arcsin(s_d/v)\bfone_{\emptyset}(-\theta_{max}\leq \zeta\leq\theta_{max}),\cr
(\pi\mp\arcsin(s_d/v))\bfone_{\emptyset}(-\theta_{max}\leq \zeta\leq\theta_{max}).}\label{s_d1}
\eqn

When the sensor observes the line on which the edge of direction $\xi$ exists from $t_s$ to $t_e$ with $\theta^*=\zeta$, the length on this line between the detected point at $t_s$ and that at $t_e$ is $|x^*(t_e)-x^*(t_s)|/|\cos\xi|=(t_e-t_s)|v\cos(\phi-\xi)|$ (Fig. \ref{basic}).
When the detected points at $t_s$ and $t_e$ are two end points of an edge of length $\lambda$ and direction $\xi$, this length on this line is $\lambda$.
Therefore,
\bq
\lambda=l_d|v\cos(\phi-\xi)|\label{lam1b}
\eq
where $l_d=t_e-t_s$ is the length in time taken to detect this edge.
Thus, due to Eq. (\ref{s_d1}) and the fact that  $\xi-\phi$ must satisfy Eq. (\ref{vert-d}),
\bq
\lambda=l_d v\sqrt{1-(s_d/v)^2}\bfone_{\emptyset}(-\theta_{max}\leq \zeta\leq\theta_{max}).\label{lam1}
\eq

\subsubsection{For $\theta^*=\pm\theta_{max}$}
When the detecting direction is $\theta^*=\pm\theta_{max}$, the line the direction of which is the same as the detecting direction and that passes through the sensor's location is 
\bq
y=(\tan(\phi\pm\theta_{max}))(x-tv\cos\phi-x_s(0))+tv\sin\phi+y_s(0).
\eq
The intersection $(x^*,y^*)$ of this line and the line defined by Eq. (\ref{line_xi}) is
\bqn
x^*(t)&=&\frac{1}{\tan(\phi\pm\theta_{max})-\tan\xi}\cr
&&\{(\tan(\phi\pm\theta_{max}))(tv\cos\phi+x_s(0))\cr
&&-tv\sin\phi-y_s(0)\},\cr
y^*(t)&=&(\tan\xi)x^*(t).
\eqn
Because $r(t)=|\triangle x(t)/\cos(\phi\pm\theta_{max})|=|(x^*(t)-tv\cos\phi-x_s(0))/\cos(\phi\pm\theta_{max})|$ when the intersection is on the edge, $r(t)$ is a linear function of $t$ and its slope $s_d$ is 
\bq
s_d= \pm\frac{v\sin(\xi-\phi)}{\sin(\phi\pm\theta_{max}-\xi)}.
\eq
Because $\xi-\phi$ must satisfy Eq. (\ref{max-d}),
\bqn
&&\xi-\phi\cr
&=&
(\arctan\frac{s_d\sin\theta_{max}}{s_d\cos\theta_{max}\pm v}+(0\,{\rm or}\, \pi))\bfone_\emptyset(\theta_{max}<\zeta),\cr
&&\xi-\phi\cr
&=&
(-\arctan\frac{s_d\sin\theta_{max}}{s_d\cos\theta_{max}\pm v}+(0\,{\rm or}\, \pi))\bfone_\emptyset(\zeta<-\theta_{max}).\label{s_d2}\cr
&&
\eqn
Similarly to the derivation of Eq. (\ref{lam1b}),
\bq
\lambda=l_d|\frac{v\sin\theta_{max}}{\sin(\phi\pm\theta_{max}-\xi)}|.
\eq
Due to Eq. (\ref{s_d2}), $\lambda=\frac{l_d|s_d|\sin\theta_{max}}{|\sin(\xi-\phi)|}$ and $\sin(\xi-\phi)=\frac{s_d\sin\theta_{max}}{\sqrt{(s_d\sin\theta_{max})^2+(s_d\cos\theta_{max}\pm v)^2}}$.
Thus,
\bqn
\lambda&=&l_d\sqrt{(s_d\sin\theta_{max})^2+(s_d\cos\theta_{max}\pm v)^2}\cr
&&\bfone_\emptyset((\theta_{max}<\zeta)\cup(\zeta<-\theta_{max})).\label{lam2}
\eqn

\subsection{Shape of $r(t)$}\label{r-shape}
\begin{figure}[tb] 
\begin{center} 
\includegraphics[width=10cm,clip]{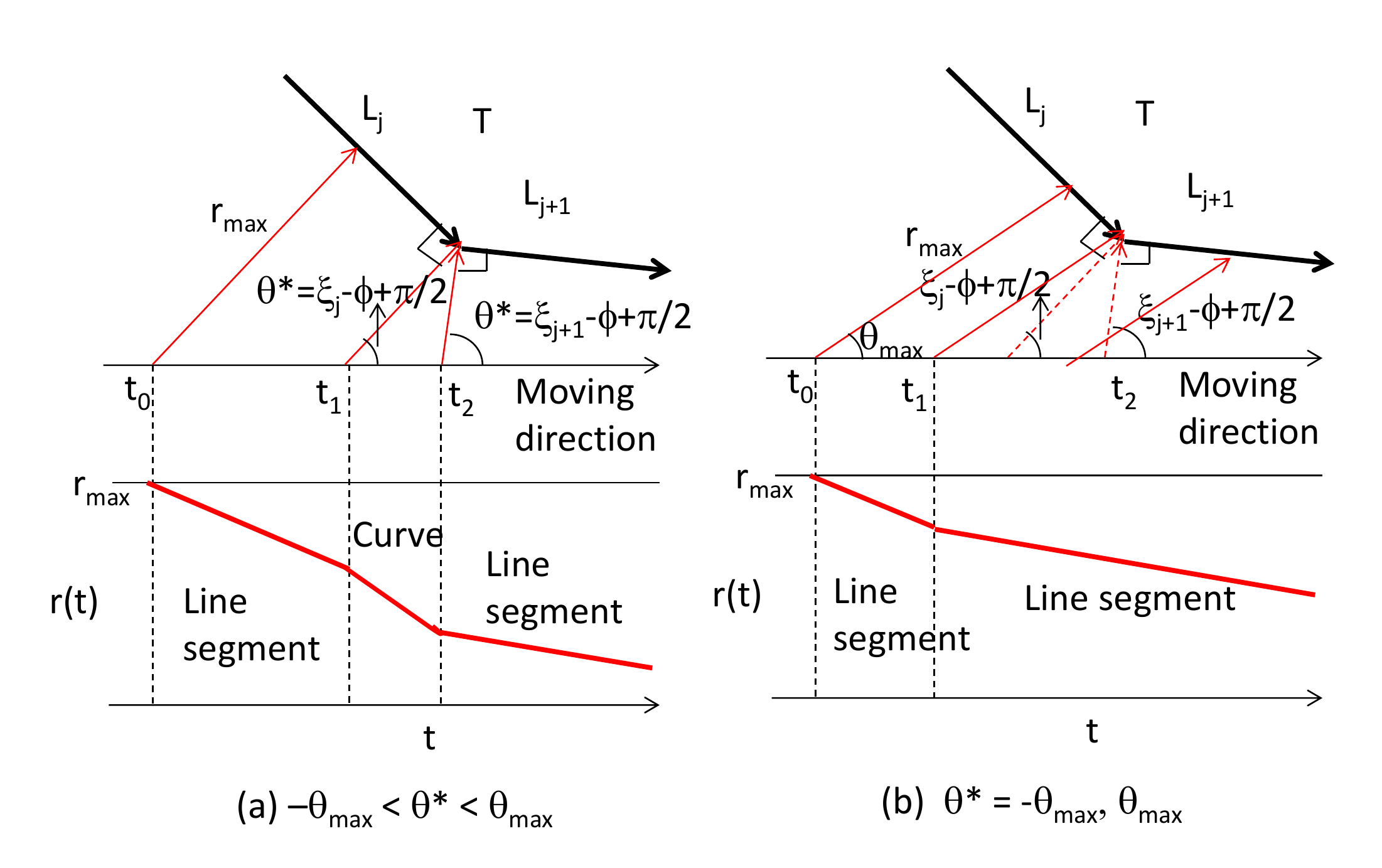} 
\caption{Illustration of $r(t)$} 
\label{curve} 
\end{center} 
\end{figure}

A sensor keeps detecting an edge $L$, $r(t)$ becomes continuous and a line segment.
At a vertex, a detecting direction changes and $r(t)$ may become a curve.
A curve appears  between $t_1$ and $t_2$ when $\theta^*=\xi_j-\phi+\pi/2\in [-\theta_{max},\theta_{max}]$ just before a vertex (Fig. \ref{curve}-(a)).
This is because the detected point is at the vertex while the sensor is moving between $t_1$ and $t_2$ and because the distance between the vertex and the sensor is not a linear function of $t$ between $t_1$ and $t_2$.
When $\theta^*=\pm\theta_{max}$ just before a vertex, no curve appears because the detected point is moving as the sensor is moving (Fig. \ref{curve}-(b)).

In the remainder of this paper, we focus on the sensing results $r(t)>0$.
When $r(t)$ becomes a line segment during a period detecting the {\it whole} $L_{j}$ with $r(t)>0$ by a sensor and the period starts at $t_s$ and ends at $t_e$, an event corresponding to $t_s$ is (i) a change of slope at $r(t_s)>0$ (a curve may end at $t_s$) or (ii) $r(t_s)<r_{max}$ and $r(t_s-dt)=\emptyset$ and an event corresponding to $t_e$ is (i) a change of slope at $r(t_e)>0$ (a curve may start at $t_e$) or (ii) $r(t_e-dt)<r_{max}$ and $r(t_e)=\emptyset$.
Here, $0<dt\ll 1$ and $t-dt$ means ^^ ^^ just before $t$."
(Note that the period does not include $r(t)=0$.)
Let $p_d(L)$ be a period of $r(t)$ starting and ending with the above-mentioned events of $L$ with $r(t)>0$ and let $l_d(L)$ be the length in time of $p_d(L)$.
We also define $s_d(L)$: $s_d(L)$ is the slope of $r(t)$ detecting $L$.
We can obtain $s_d$ if we cannot observe the whole $L$ but observe a partial $L$ through a single sensor.
Therefore, to obtain $s_d$, the start epoch $t_s$ can be $r(t_s)=0,r_{max}$ in addition to (i)-(ii) for $r(t_s)$ mentioned above and the end epoch $t_e$ can be $r(t_e)=0,r_{max}$ in addition to (i)-(ii) for $r(t_e)$ mentioned above.

\begin{remark}\label{remark1}
$s_d(L_j)$ for the period with $r(t_e)=0$ and $s_d(L_{j+1})$ for the period with $r(t_s)=0$ are not useful for the following reason. As proposed in a later section, to estimate an angle $\gamma_j$, we need $s_d(L_j)$ and $s_d(L_{j+1})$, i.e., $s_d$ for consecutive edges. This is because we derive the estimate of $\gamma_j$ from the estimate of $\xi_j-\phi$ and that of $\xi_{j+1}-\phi$. If there are a period ending at $t_e$ with $r(t_e-dt)>0$, another period of $r(t)=0$ for $\forall t\in [t_e, t_s']$, and a period starting at $t_s'$ with $r(t_s'+dt)>0$, we cannot estimate an angle from $s_d$ for the combination of the first and third periods. This is because these two pieces of $s_d$ may not be $s_d$ for consecutive edges. 
\end{remark}

\subsection{Probability that sensor detects whole $L$}\label{sec-p_d}
\begin{figure}[tb] 
\begin{center} 
\includegraphics[width=11cm,clip]{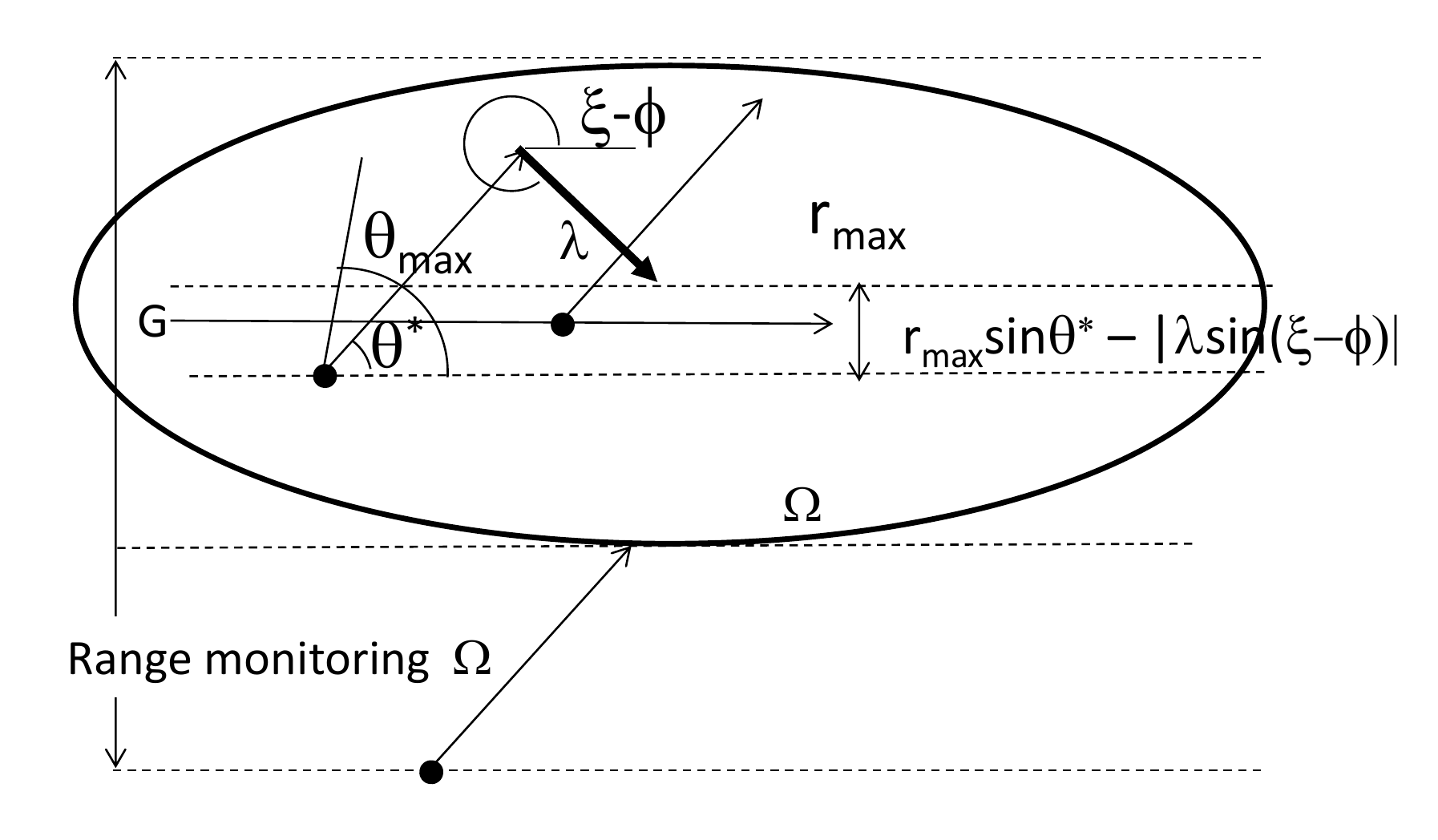} 
\caption{Location of sensors detecting whole edge with $r(t)>0$}
\label{omega2} 
\end{center} 
\end{figure}

According to Fig. \ref{omega2}, if a directional line $G$ on which the sensor moves is in the strip of width $r_{max}\sin\theta^*-\lambda|\sin(\xi-\phi)|$ and if $\phi$ satisfies Eq. (\ref{fix-theta-xi}), the sensor can detect the whole $L$.
Here, $\theta^*$ is determined by Eq. (\ref{detecting_direction}).
Because the strip width must be non-negative, $$r_{max}|\sin(\zeta)|>\lambda|\sin(\xi-\phi)|$$ for $\zeta\in[-\theta_{max},\theta_{max}]$ and $$r_{max}|\sin\theta_{max}|>\lambda|\sin(\xi-\phi)|$$ for $\zeta\in[\theta_{max},\theta_{max}+\pi/2]\cup[-\theta_{max}-\pi/2,-\theta_{max}]$.
Thus, for $\zeta\in[-\theta_{max},\theta_{max}]$, 
\begin{eqnarray*}
\phi-\xi&\in&[-\arctan\frac{r_{max}}{\lambda},\arctan\frac{r_{max}}{\lambda}]\cr
&&\cup[\pi-\arctan\frac{r_{max}}{\lambda},\pi+\arctan\frac{r_{max}}{\lambda}],
\end{eqnarray*}
and for $\zeta\in[\theta_{max},\theta_{max}+\pi/2]\cup[-\theta_{max}-\pi/2,-\theta_{max}]$, $\phi-\xi\in[-\eta,\eta]\cup[\pi-\eta,\pi+\eta]$.
Here, $\eta(\lambda)\defeq\cases{\pi/2,  & for $r_{max}|\sin\theta_{max}|\geq\lambda$,\cr
\arcsin\frac{r_{max}|\sin\theta_{max}|}{\lambda}\in [0,\pi/2], &otherwise.}$

Note that the measure of the set of $G$ on which sensors monitor $\Omega$ (Fig. \ref{omega2}) is given by Eq. (5.2) in \cite{santalo} and is $\lengthx{\Omega}+\pi r_{max}\sin\theta_{max}$.
Also note that the measure $\measure_1$ of the set of $G$ that is in this strip and has a direction satisfying  Eq. (\ref{fix-theta-xi}) is as follows where
$\Phi_{1,1}(\xi)\defeq[\zeta_m,\zeta_p]\cap([\xi-\arctan\frac{r_{max}}{\lambda},\xi+\arctan\frac{r_{max}}{\lambda}]\cup[\xi+\pi-\arctan\frac{r_{max}}{\lambda},\xi+\pi+\arctan\frac{r_{max}}{\lambda}])$ and 
$\Phi_{1,2}(\xi)\defeq([\zeta_m-\pi/2,\zeta_m]\cup[\zeta_p,\zeta_p+\pi/2])\cap([\xi-\eta,\xi+\eta]\cup[\xi+\pi-\eta,\xi+\pi+\eta])$.
\bqn
&&\measure_1(\lambda)\cr
&=&\int_{\Phi_{1,1}(\xi)}r_{max}|\sin(\zeta)|-\lambda|\sin(\xi-\phi)|d\phi\cr
&+&\int_{\Phi_{1,2}(\xi)}r_{max}|\sin\theta_{max}|-\lambda|\sin(\xi-\phi)|d\phi\cr
&=&\bfone(\frac{\pi}{2}-\theta_{max}<\arctan\frac{r_{max}}{\lambda})2\int_{\pi/2-\theta_{max}}^{\arctan\frac{r_{max}}{\lambda}}r_{max}\cos x -\lambda|\sin x|dx\cr
&+&\bfone(\eta\geq\max(\pi/2-\theta_{max},\theta_{max}))2\int_{-\theta_{max}}^{\pi/2-\theta_{max}}r_{max}\sin\theta_{max}-\lambda|\sin x|dx\cr
&+&\bfone(\pi/2-\theta_{max}\leq\eta<\theta_{max})2\int_{-\eta}^{\pi/2-\theta_{max}}r_{max}\sin\theta_{max}-\lambda|\sin x|dx\cr
&+&\bfone(\theta_{max}\leq\eta<\pi/2-\theta_{max})2\int_{-\theta_{max}}^{\eta}r_{max}\sin\theta_{max}-\lambda|\sin x|dx\cr
&+&\bfone(\eta<\min(\pi/2-\theta_{max},\theta_{max}))2\int_{-\eta}^{\eta}r_{max}\sin\theta_{max}-\lambda|\sin x|dx\cr
&=&\bfone(\frac{\pi}{2}-\theta_{max}<\arctan\frac{r_{max}}{\lambda})\cr
&&\qquad 2\{r_{max}(\frac{r_{max}}{\sqrt{\lambda^2+r_{max}^2}}-\cos\theta_{max})+\lambda(\frac{\lambda}{\sqrt{\lambda^2+r_{max}^2}}-\sin\theta_{max})\}\cr
&+&\bfone(\eta\geq\max(\pi/2-\theta_{max},\theta_{max}))\cr
&&\qquad2\{(\pi/2) r_{max}\sin\theta_{max}-\lambda(2-\cos\theta_{max}-\sin\theta_{max})\}\cr
&+&\bfone(\pi/2-\theta_{max}\leq\eta<\theta_{max})\cr
&&\qquad2\{(\pi/2-\theta_{max}+\eta) r_{max}\sin\theta_{max}-\lambda(2-\cos\eta-\sin\theta_{max})\}\cr
&+&\bfone(\theta_{max}\leq\eta<\pi/2-\theta_{max})\cr
&&\qquad2\{(\eta+\theta_{max})r_{max}\sin\theta_{max}-\lambda(2-\cos\theta_{max}-\cos\eta)\}\cr
&+&\bfone(\eta<\min(\pi/2-\theta_{max},\theta_{max}))4\{\eta r_{max}\sin\theta_{max}-\lambda(1-\cos\eta)\},\label{measure_1}
\eqn
Because the probability $q_d(\lambda)$ that the sensor detects the whole $L$ of length $\lambda$ is given by the ratio of these measures in accordance with the definition of geometric probability \cite{santalo}, 
\bq
q_d(\lambda)=\frac{\measure_1(\lambda)}{2(\lengthx{\Omega}+2\pi r_{max} \sin\theta_{max})}\label{q_d_lam}
\eq
(The denominator doubles because $G$ is directional.)
Therefore, the expected number $E[n_d(\lambda)]$ of sensors detecting the whole $L$ of length $\lambda$ with $r(t)>0$ is given by 
\bq
E[n_d(\lambda)]=n_sq_d(\lambda).\label{num_detects1}
\eq

\subsection{Probability that sensor detects a vertex}
Here, we pay attention to the number of sensors that have sensing results that cover a vertex of $T$.
Such sensing results may not cover a whole edge.

\begin{figure}[tb] 
\begin{center} 
\includegraphics[width=11cm,clip]{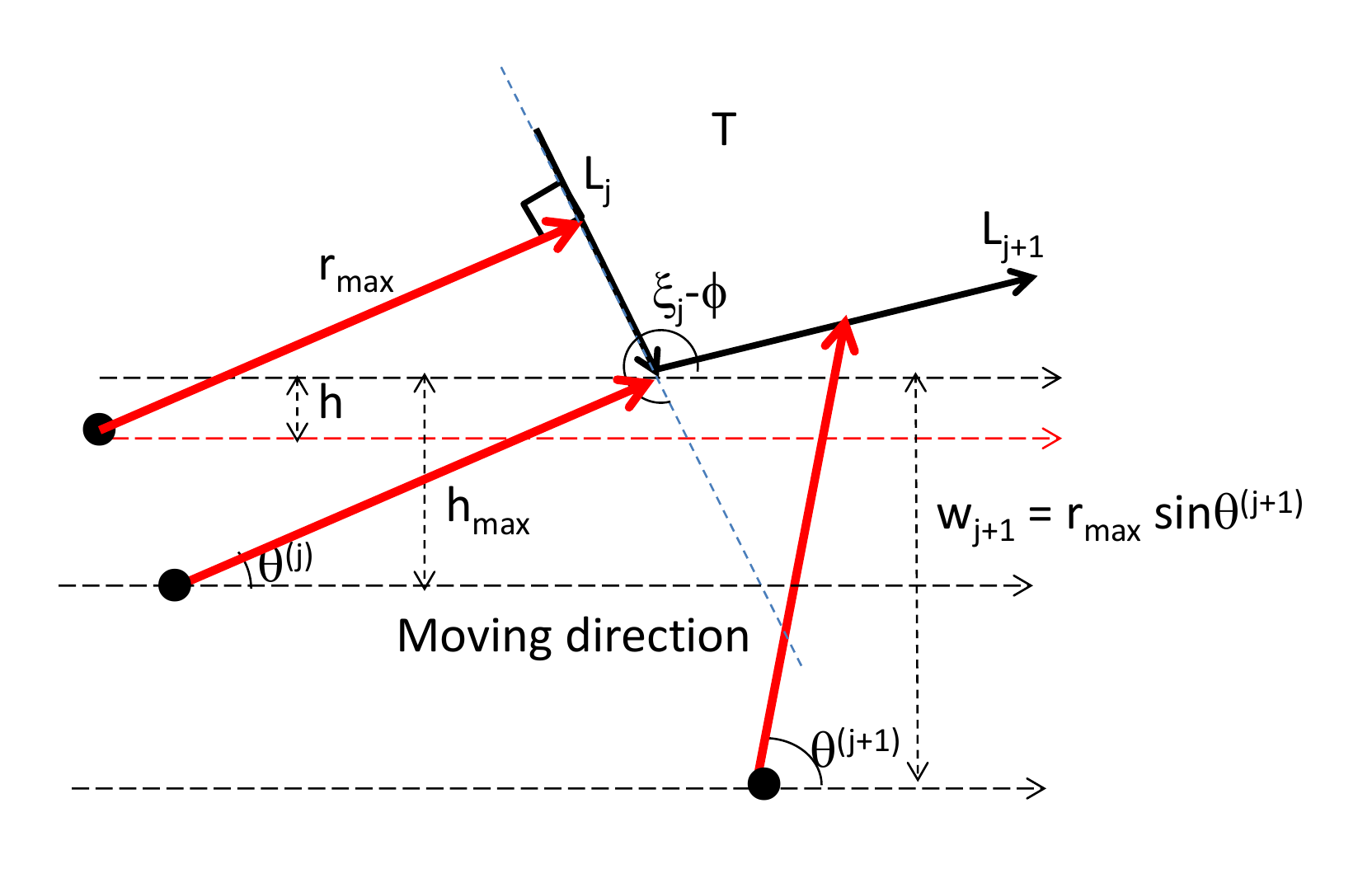} 
\caption{Location of sensors detecting whole edge with $r(t)>0$}
\label{single-edge} 
\end{center} 
\end{figure}

Focus on a vertex formed by $L_j,L_{j+1}$ and assume that the vertex is on the left-side of line $G$ on which a sensor is moving.
Because the vertex on the left-side of line $G$ is detected, $0<\theta^{(k)}<\pi$ for $k=j,j+1$.
Here, $\theta^{(k)}$ is the detecting direction for $L_k$.
Thus, due to Eq. (\ref{detecting_direction}), $\{\xi_k-\theta_{max}<\phi<\xi_k-\theta_{max}+\pi/2\}\cup(\{0<\xi_k+\pi/2-\phi<\pi\}\cap\{\xi_k-\theta_{max}+\pi/2<\phi<\xi_k+\theta_{max}+\pi/2\})$.
This means that $\xi_k-\theta_{max}<\phi<\xi_k+\pi/2$.
To detect $L_j,L_{j+1}$ around the vertex, 
\bqn
\phi\in\Phi_2
&\defeq&(\xi_j-\theta_{max},\xi_j+\pi/2)\cap(\xi_{j+1}-\theta_{max},\xi_{j+1}+\pi/2).\label{Phi_2}
\eqn
In addition, we need a condition in which the detected point on $L_j$ is not always at the vertex. 
This condition is equivalent to (i) $\theta^{(j)}=\theta_{max}$ or (ii) $0<\theta^{(j)}=\xi_j+\pi/2-\phi<\pi, \xi_j-\theta_{max}+\pi/2<\phi<\xi_j+\theta_{max}+\pi/2$ and the detected point is on $L_j$ not equal to the vertex formed by $L_j,L_{j+1}$.
Otherwise, we do not obtain any information on $s_d(L_j)$.

This (i) means $\theta_{max}-\pi/2\leq\xi_j-\phi<\theta_{max}$, and (ii) means $-\pi/2\leq\xi_j-\phi<-\pi/2+\theta_{max}$ (equivalently, $3\pi/2\leq\xi_j-\phi<3\pi/2+\theta_{max}$) and $h<h_{max}$ (Fig. \ref{single-edge}).
Here, $h_{max}(\phi,\xi_j)\defeq r_{max}\sin(\xi_j-\phi+\pi/2)$.
For simplicity, define $h_{max}=\infty$ for $\theta_{max}-\pi/2\leq\xi_j-\phi<\theta_{max}$ and 0 for $\theta_{max}\leq\xi_j-\phi<3\pi/2$.
According to Fig. \ref{single-edge}, the measure $\measure_2$ of the set of $G$ on which a sensor detects a part around the vertex is 
\bqn
\measure_2&=&2\int_{\Phi_2}\min(w_j,w_{j+1},h_{max}(\phi,\xi_j))d\phi.\label{measure_2}
\eqn
Here, 2 on the right-hand side of the equation above appears due to the symmetry of the assumption ^^ ^^ the vertex is on the left-side of line $G$", and $w_k=r_{max}\sin(\xi_k+\pi/2-\phi)$ for $\phi\in[\xi_k+\pi/2-\theta_{max},\xi_k+\pi/2+\theta_{max}]$ and $w_k=r_{max}\sin\theta_{max}$ otherwise.

It is possible to calculate $\measure_2$ like $\measure_1$ although it is omitted.
Numerical integration based on Eq. (\ref{measure_2}) is also possible to obtain $\measure_2$.

\begin{remark}
$\measure_2$ is a function of $\xi_j$ and $\xi_{j+1}$.
However, it is a function of $\gamma_j$ because we can use any direction as a reference direction.
\end{remark}

Similar to in Subsection \ref{sec-p_d}, the probability $q_d(\gamma_j)$ that the sensor can detect the vertex formed by $L_j,L_{j+1}$ with $r(t)>0$ is given by the following.
\bq
q_d(\gamma_j)=\frac{\measure_2}{2(\lengthx{\Omega}+2\pi r_{max} \sin\theta_{max})}.
\eq

Therefore, the expected number $E[n_d(\gamma)]$ of sensors detecting a vertex of inner angle $\gamma$ with $r(t)>0$ is given by 
\bq
E[n_d(\gamma)]=n_s q_d(\gamma). \label{num_vertex}
\eq

\begin{remark}
Due to Eq. (\ref{Phi_2}), $\phi$ must be in $\Phi_2=[\max(\xi_{j+1},\xi_j)-\theta_{max},\min(\xi_{j+1},\xi_j)+\pi/2]$ to detect a vertex formed by $L_j,L_{j+1}$.
Thus, the length of the range of $\phi$ satisfying $\Phi_2$ is $[\gamma_j-\pi/2+\theta_{max}]^+$.
Therefore, we cannot detect a vertex of angle $\gamma<\pi/2-\theta_{max}$.
\end{remark}

\section{Estimation method}
\subsection{Estimating edge lengths and angles of $T$}
According to the definition of starting and ending events on $r(t)$ defined in Subsection \ref{r-shape}, we obtain $s_d$ or $l_d$ for each sensor detecting $T$.
By using them, we estimate the shape of $T$.

Edge length $\lambda$ is estimated through Eqs. (\ref{lam1}) and (\ref{lam2}) and angle $\gamma$ through Eqs. (\ref{s_d1}) and (\ref{s_d2}). For the $\lambda$ estimation, applying Eqs. (\ref{lam1}) and (\ref{lam2}) to $l_d$ (and $s_d$) can directly yield three estimates of $\lambda$.
\bqn
\widehat{\lambda}&=&l_d v\sqrt{1-(s_d/v)^2}\\
\widehat{\lambda}&=&l_d\sqrt{(s_d\sin\theta_{max})^2+(s_d\cos\theta_{max}\pm v)^2}\label{hat_lam2}
\eqn

On the other hand, Eqs. (\ref{s_d1}) and (\ref{s_d2}) directly estimate $\xi-\phi$ but not $\gamma$. On the basis of the estimates of $\xi-\phi$, we derive the estimate of $\gamma$. Assume that a sensor detects $L_j$ and $L_{j+1}$ and that the slopes of $r(t)$ for them are $s_d^{(j)}$ and $s_d^{(j+1)}$. Because $\gamma_j=\pi-\xi_{j+1}+\xi_j$, we obtain the estimates of $\gamma_j$.
\bq
\widehat{\gamma_j}=\pi-\widehat{\xi_{j+1}-\phi}+\widehat{\xi_{j}-\phi}\label{hat_gamma}
\eq
Here, $\widehat{\xi_{k}-\phi}$ is given by Eq. (\ref{s_d1}) or (\ref{s_d2}) and $s_d^{(k)}$.

If we exactly obtain $s_d$ and $l_d$ (and its associated period determined by $t_s$ and $t_e$), the estimates mentioned above are exact.
However, we cannot uniquely obtain estimate $\lambda$ or $\gamma$.
This is because $\theta^*$ is unknown and because Eqs. (\ref{s_d1}), (\ref{s_d2}), and (\ref{hat_gamma}) cannot uniquely determine $\widehat{\xi_{k}-\phi}$ or $\widehat{\lambda}$.
We need to overcome this non-uniqueness.

In our proposed method, we choose an estimate from multiple estimates as follows. For each pair of $(l_d,s_d)$ or each pair of $(s_d^{(j)},s_d^{(j+1)})$, we can obtain multiple $\widehat{\lambda}$ or $\widehat{\gamma_j}$. Call these multiple $\widehat{\lambda}$ or $\widehat{\gamma_j}$ candidate estimates of $\lambda$ or $\gamma_j$. Let $\estcan_set(l_d,s_d)$ ($\estcan_set(s_d^{(j)},s_d^{(j+1)})$) be the set of these candidate estimates derived from $(l_d,s_d)$($(s_d^{(j)},s_d^{(j+1)})$). The set of these candidate estimates includes at least one exact estimate because the estimates mentioned above do not include errors. When the number of $(l_d,s_d)$-samples ($(s_d^{(j)},s_d^{(j+1)})$-samples) is large, the number of occurrences of exact estimates is also large. For example, if $k$ $(l_d,s_d)$-samples are detecting an edge of length $\lambda$, the number of occurrences of $\widehat{\lambda}=\lambda$ is larger than or equal to $k$. On the other hand, the number of occurrences of other candidate estimates is less than the exact one because other candidate estimates depend on $(l_d,s_d)$ or $(s_d^{(j)},s_d^{(j+1)})$. Figure \ref{simple_default} clarifies this. Additional comments are provided in Appendix. Thus, by counting the numbers of occurrences, we adopt the candidate estimates that occur many more times than others as estimates.

Specifically, the proposed method chooses an estimate among $\cup_{(l_d,s_d)}\estcan_set(l_d,s_d)$ ($\cup_{(s_d^{(j)},s_d^{(j+1)})}\estcan_set(s_d^{(j)},s_d^{(j+1)})$) as follows.
Equally divide the interval between the smallest candidate estimate $\min\{\widehat{\lambda}\in\cup_{(l_d,s_d)}\estcan_set(l_d,s_d)\}$ and the largest candidate estimates $\max\{\widehat{\lambda}\in\cup_{(l_d,s_d)}\estcan_set(l_d,s_d)\}$ (the smallest $\min\{\widehat{\gamma}\in\cup_{(s_d^{(j)},s_d^{(j+1)})}\estcan_set(s_d^{(j)},s_d^{(j+1)})\}$ and the largest  $\max\{\widehat{\gamma}\in\cup_{(s_d^{(j)},s_d^{(j+1)}}\estcan_set(s_d^{(j)},s_d^{(j+1)})\}$) into $n_{sub}$ sub-intervals.
Determine $n_{sub}$ such that the peak of the number of occurrences of candidate estimates is clear (Figure \ref{simple_default}).
Typically, $k_{sub}$ (the ratio of the total number of candidate estimates to the number of sub-intervals) is several such as five or ten.
Formally, $k_{sub}$ is defined as $\sharp(\cup_{(l_d,s_d)}\estcan_set(l_d,s_d))/n_{sub}$ or $\sharp(\cup_{(s_d^{(j)},s_d^{(j+1)})}\estcan_set(s_d^{(j)},s_d^{(j+1)}))/n_{sub}$.

Then, count the number $c(\widehat{\lambda})$ ($c(\widehat{\gamma})$) of occurrences of candidate estimates in a sub-interval where the average of candidate estimates in this sub-interval is $\widehat{\lambda}$ $(\widehat{\gamma})$.
If the count in a certain sub-interval is larger than a threshold, calculate $\widehat{\lambda}$ $(\widehat{\gamma})$, which is the average of candidate estimates in that sub-interval, and adopt it as an estimate.
In the remainder of this section, we focus on the adopted estimate.

\begin{remark}
If some errors, typically sensing errors, are likely contained, the number of sub-intervals should decrease (equivalently, the sub-interval length should be longer) to merge similar candidates. This is because there are many candidate estimates around the exact length or angle under some errors. To use the method to estimate the number of edges and vertexes described in the next subsection, these candidates should be merged.
\end{remark}

\subsection{Estimating the number of edges and vertexes}
We need to estimate an edge length $\lambda$,  a vertex angle $\gamma$, the number of the edges of length $\lambda$, and the number of the vertexes of angle $\gamma$.
The previous subsection covers the estimation method for $\lambda$ and $\gamma$.
This subsection covers the latter two estimations on the basis of Eqs. (\ref{num_detects1}) and (\ref{num_vertex}).

Let $n_\lambda$ be the number of whole edge detection samples and $n_\gamma$ be the number of vertex detection samples. $n_\lambda$ is equal to the number of samples of $l_d>0$, and $n_\gamma$ be the number of $s_d$ pairs of consecutive vertexes. Note that $c(\widehat{x})/\sum_{\widehat{y}\in\est_set(x)} c(\widehat{y})$ for $x=\lambda,\gamma$ is the ratio of the number of occurrences of estimates $\widehat{x}$ among the total number of occurrences of estimates. Here, $\est_set(\lambda)$ ($\est_set(\gamma)$) means the set of estimates of edge length (angle). Thus, the mean number of whole edge detection samples for edge length $\widehat{\lambda}$ (or vertex detection samples for angle $\widehat{\gamma}$) is $n_xc(\widehat{x})/\sum_{\widehat{y}\in\est_set(x)} c(\widehat{y})$. Then, the estimated number $\widehat{N_{\lambda}}$ of edges of length $\lambda$ and the estimated number $\widehat{N_{\gamma}}$ of vertexes of angle $\gamma$ are 
\bqn
\widehat{N_{\widehat{x}}}&=&\frac{n_xc(\widehat{x})}{E[n_d(\widehat{x})]\sum_{\widehat{y}\in\est_set(x)} c(\widehat{y})}.\label{N_x}
\eqn
After deriving edge length estimates $\widehat{\lambda}$ and angle estimates $\widehat{\gamma}$, evaluate Eqs. (\ref{num_detects1}) and (\ref{num_vertex}), obtain $E[n_d(\widehat{\lambda})]$ and $E[n_d(\widehat{\gamma})]$, and use them in Eq. (\ref{N_x}) to derive $\widehat{N_{\widehat{x}}}$.

\subsection{Estimating the shape of $T$}
Even when we estimate the length of each edge and the angle of each vertex, we cannot identify the shape of $T$.
We need to know the sequence of edges or vertexes.
To identify the sequence, use the following method.

Assume that there exist sensing results $l_d(L_j),s_d^{(j)},s_d^{(j+1)}$ for a single sensor. That is, there is a sensor that detects the whole $L_j$ and a part of $L_{j+1}$ including a vertex formed by $L_j$ and $L_{j+1}$. (As mentioned in Remark \ref{remark1}, there should not be a period of $r(t)=0$ between the period of $p_d(L_j)$ and that detecting $L_{j+1}$.) Through the sensing results $l_d(L_j),s_d^{(j)}$, we obtain the estimate $\widehat{\lambda_j}$. We can also obtain the estimate $\widehat{\gamma_j}$ through $s_d^{(j)},s_d^{(j+1)}$. This means that one vertex of an edge of length $\widehat{\lambda_j}$ has an angle $\widehat{\gamma_j}$. Note that we cannot identify $j$. Thus, this means that we know there is an edge of length $\widehat{\lambda}$ connected at a vertex of angle $\widehat{\gamma}$. Let $\lambda[k]$ and $\gamma[k]$ be such a pair of an edge length estimate and an angle estimate the edge connects where they were derived by the $k$-th sensor.

By using many sensing results, we obtain $\{\lambda[k],\gamma[k]\}_k$. Let $\sensor_id(a,b)$ be the set of sensors (sensor identifiers) $\{k|\lambda[k]=a\in\est_set(\lambda), \gamma[k]=b\in\est_set(\gamma) \}$. If an angle estimate and an edge length estimate are independent, the expected number $E_{ind}[\sharp\sensor_id(a,b)]$ of the elements in $\sensor_id(a,b)$ is 
\bqn
E_{ind}[\sharp\sensor_id(a,b)]
&=&\frac{E[N_d(a)]E[N_d(b)]\sum_{\widehat{\lambda}\in\est_set(\lambda),\widehat{\gamma}\in\est_set(\gamma)}\sharp\sensor_id(\widehat{\lambda},\widehat{\gamma})}{\sum_{\widehat{\lambda}\in\est_set(\lambda),\widehat{\gamma}\in\est_set(\gamma)}E[N_d(\widehat{\lambda})]E[N_d(\widehat{\gamma})]}.
\eqn
This is because $E[N_d(\lambda)]\defeq \widehat{N_{\lambda}}E[n_d(\lambda)]$ is the expected number of whole edge detections for any edge of length $\lambda$ and $E[N_d(\gamma)]\defeq \widehat{N_{\gamma}}E[n_d(\gamma)]$ is the expected number of vertex detections for any vertex of angle $\gamma$. If, however, the observed number of elements in this set is much larger (less) than this theoretical value, an edge of length $a$ connecting at a vertex of angle $b$ is likely (unlikely) to exist.

By finding the set of pairs of an edge length and a vertex angle connected by the edge, we can make a table such as Table \ref{comb_table1}. Such a table enables us to guess that an edge of a certain length connects two vertexes of certain angles or that a vertex of a certain angle is formed by two edges of certain lengths. Then, by sequentially connecting them, we can identify the shape of $T$. For example, if such a table suggests that there are a single vertex (A) formed by edges (a) and (b), a single vertex (B) formed by edges (b) and (c), and a single vertex (C) formed by edges (c) and (a), we can estimate that $T$ is a triangle of vertexes (A), (B), and (C) and that the edge between (A) and (B) ((B) and (C); (C) and (A)) is (b) ((c) and (a)).

\begin{remark}\label{shape_remark}
More precisely, we may not able to identify the shape of $T$ or sequentially connect edges or vertexes even when two vertex angles of each edge are given or two edge lengths forming each vertex are given. For example, when all the vertex angles are the same, the shape of $T$ is difficult to identify. For two long edges connecting vertexes of $\pi/2$ and two short edges connecting vertexes of $\pi/2$, $T$ may be a rectangle. However, the two short edges may be consecutive, and the two long edges may also be consecutive. If so, the shape of $T$ becomes unnatural because we cannot obtain a close boundary of $T$. However, we cannot conclude that a non-close boundary is unnatural because even the rectangle $T$ may not have a close boundary due to estimation errors. 
\end{remark}

\section{Numerical examples}
\subsection{Default conditions}
Unless explicitly mentioned otherwise, numerical examples in this paper use the following conditions. $\Omega$ is a disk with radius 100. $T$ is placed near the center of $\Omega$ to remove the boundary effect. $r_{max}=100$, $\theta_{max}=\pi/2$, $v=0.1$, $n_s=1000$. Sensing areas of all the sensors intersect $\Omega$ but may not detect $T$. In the simulation, we move vehicles at each time unit, that is, a discrete time, not a continuous time. As a result, observed parameters such as $l_d$ cannot take a continuous value and cause some observation errors. This may result in estimation errors.

To understand the proposed estimation framework, we provide a simple figure as $T$ as a default.
(Realistic $T$s are used later.)
The simple $T$ is a right triangle of edge lengths of 50, 25, and $25\sqrt{3}$.

\subsection{Estimation under default conditions}
\begin{figure}[tb] 
\begin{center} 
\includegraphics[width=11cm,clip]{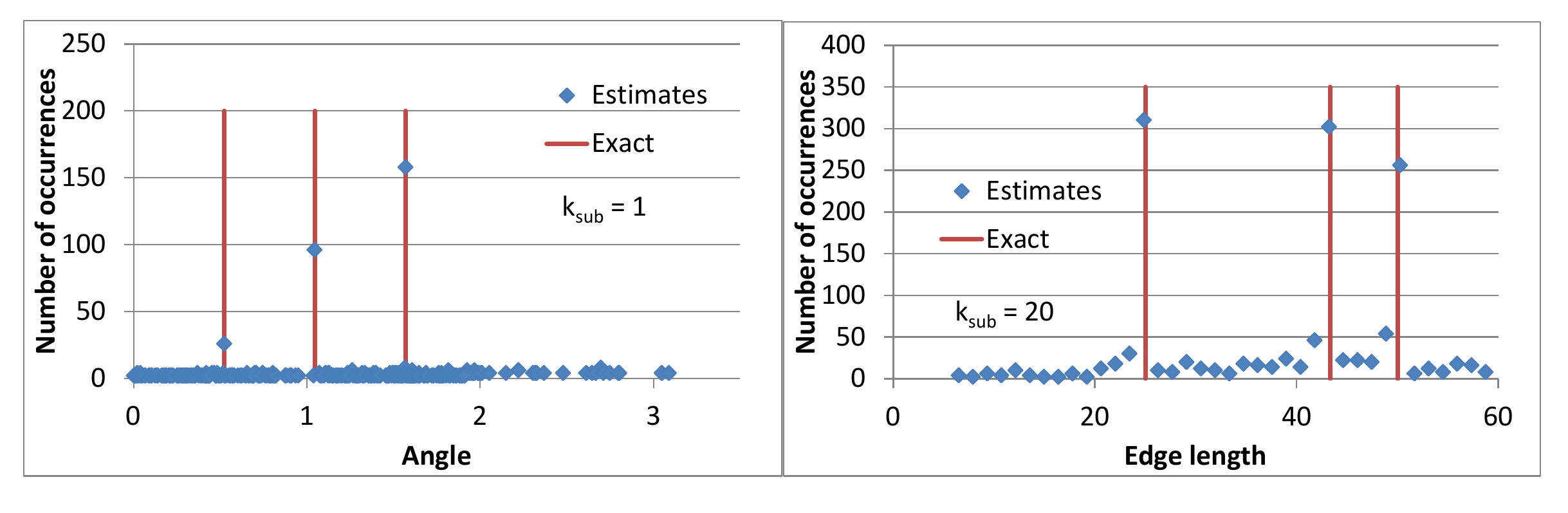} 
\caption{Estimated angles and lengths of simple target object under default conditions}
\label{simple_default} 
\end{center} 
\end{figure}

Figure \ref{simple_default} plots the number of candidate estimates in each sub-interval for a edge length or an angle under the default conditions. Table \ref{result_table1} summarizes the estimated lengths, angles, and their estimated numbers. Here, ^^ ^^ Estimation error" is defined by $100(\widehat{\gamma}/\gamma-1)$ or $100(\widehat{\lambda}/\lambda-1)$ where we consider that $\widehat{\gamma}$ ($\widehat{\lambda}$) is the estimate of $\gamma$ ($\lambda$), which is the closest to $\widehat{\gamma}$ ($\widehat{\lambda}$).

As shown clearly in Fig. \ref{simple_default}, the proposed method can estimate the edge lengths and angles. The angle estimates were more accurate than the length estimates (Table \ref{result_table1}). This seems to be because the discrete time sampling affects edge-lengths more than angles. 

$\widehat{N_\lambda}$ and $\widehat{N_\gamma}$ were less accurate than $\widehat{\lambda}$ and $\widehat{\gamma}$.
In particular, $\widehat{N_\gamma}$ for vertexes (B) and (C) in Table \ref{result_table1} was overestimated by more than 20\%.
This is because the estimation for the lengths and angles in the proposed framework does not include errors if we can correctly sample data and judge the whole edge detection and vertex detection.
However, $\widehat{N_\lambda}$ and $\widehat{N_\gamma}$ uses the comparison between the expected number of the whole edge detection (vertex detection) and its sample number, and the sample number is a random variable and can include errors.
Thus, $\widehat{N_\lambda}$ and $\widehat{N_\gamma}$ can become inaccurate and their accuracy seems to depend on the number of samples or the number of sensors.

\begin{table}
\caption{Summary of estimated results for a triangle $T$ under default conditions}
\begin{center}\label{result_table1}
\begin{tabular}{rlll}
\hline
Edge ID&$\widehat{\lambda}$&Estimation error (\%)&$\widehat{N_\lambda}$\\
a&24.84&-0.6093&0.8766\\
b&43.21&-0.2064&1.001\\
c&50.28&0.5501&0.9042\\
Vertex ID&$\widehat{\gamma}$&Estimation error (\%)&$\widehat{N_\gamma}$\\
A&1.570&-0.02019&0.9840\\
B&1.046&-0.0996&1.202\\
C&0.5241&-0.08829&1.223\\
\hline
\end{tabular}
\end{center}
\end{table}

Table \ref{comb_table1} summarizes $\sharp\sensor_id(a,b)$ normalized by $E_{ind}[\sharp\sensor_id(a,b)]$.
For example, Table \ref{comb_table1} strongly suggests that edge (a) connects vertexes (A) and (B), edge (b) connects vertexes (A) and (C),  and edge (c) connects vertexes (B) and (C).
Hence, we can identify the shape of the triangle $T$.

\begin{table}
\caption{Observed $\sharp\sensor_id(a,b)/E_{ind}[\sharp\sensor_id(a,b)]$ for a triangle $T$ under default conditions}
\begin{center}\label{comb_table1}
\begin{tabular}{rlll}
\hline
Edge ID/vertex ID&A&B&C\\
a&1.535&1.282&0.5409\\
b&1.462&0.1504&1.249\\
c&0.1617&1.153&0.9825\\
\hline
\end{tabular}
\end{center}
\end{table}

\subsection{Impact of number of sensors}
\begin{figure}[tb] 
\begin{center} 
\includegraphics[width=11cm,clip]{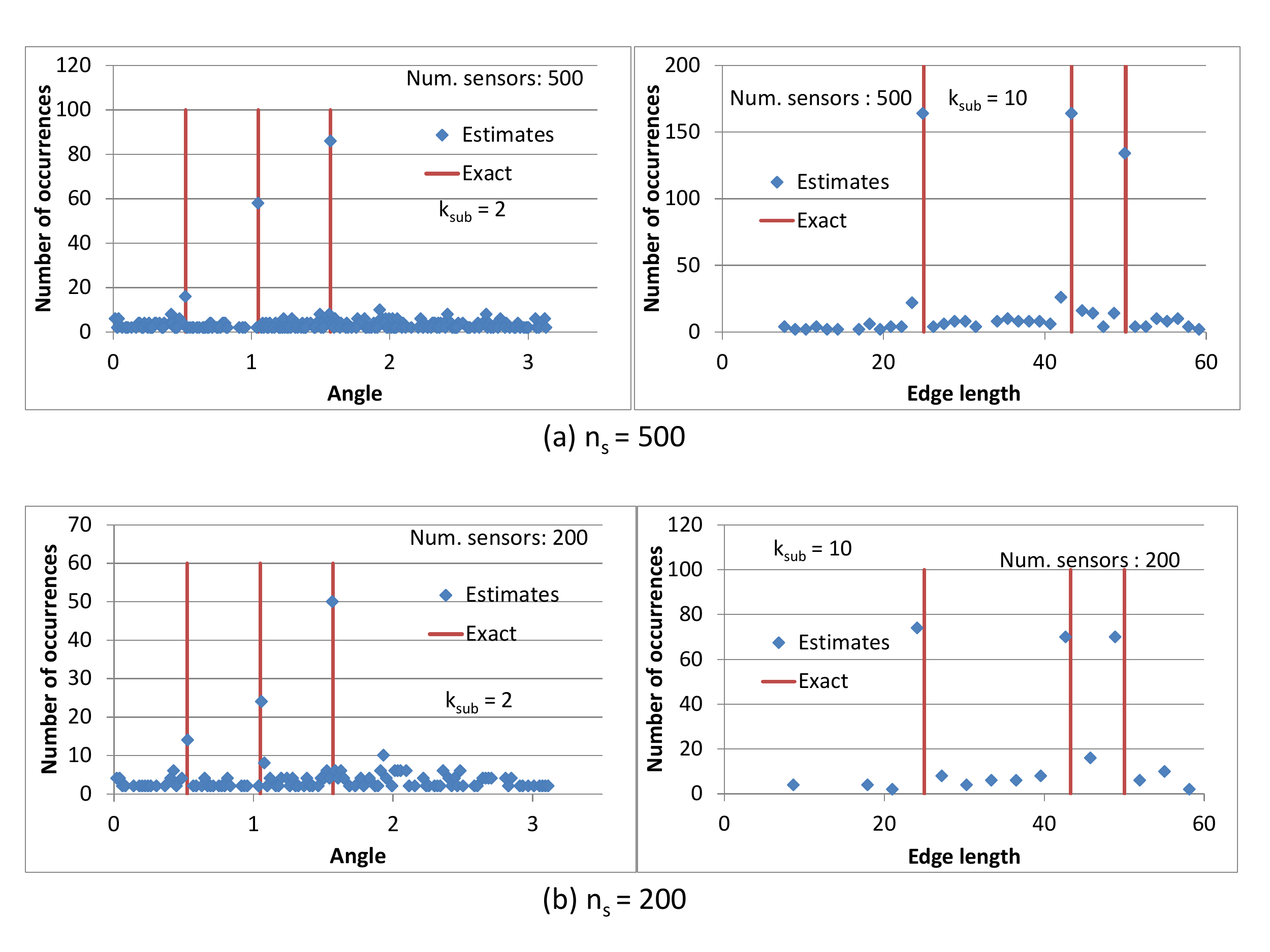} 
\caption{Estimated angles and lengths for various number of sensors}
\label{num_sensors}
\end{center} 
\end{figure}

Here, we discuss the impact of the number $n_s$ of sensors. Figure \ref{num_sensors} plots the number of estimates in each sub-interval for a edge length or an angle when 200 or 500 sensors are used. The accuracy was slightly worse than that for $n_s=1000$, but the estimates with 200 or 500 sensors are acceptable. However, $\widehat{N_\gamma}$ became inaccurate for $n_s=200$ (Table \ref{result_table_200}). $\widehat{N_\gamma}$ became nearly three for vertex (C), although it should be one. This is because the number of samples affects $\widehat{N_\gamma}$ and $\widehat{N_\lambda}$ but barely affects $\widehat{\gamma}$ and $\widehat{\lambda}$. ($\widehat{N_\lambda}$ was much more accurate than $\widehat{N_\gamma}$ because $n_\lambda \gg n_\gamma$.)

\begin{table}
\caption{Summary of estimated results for a triangle $T$ with $n_s=200$}
\begin{center}\label{result_table_200}
\begin{tabular}{rlll}
\hline
Edge ID&$\widehat{\lambda}$&Estimation error (\%)&$\widehat{N_\lambda}$\\
a&24.094&-3.624&0.9337\\
b&42.67&-1.461&1.036\\
c&48.86&-2.280&1.096\\
Vertex ID&$\widehat{\gamma}$&Estimation error (\%)&$\widehat{N_\gamma}$\\
A&1.567&-0.3183&1.407\\
B&1.057&0.9508&1.326\\
C&0.5281&0.8723&2.917\\
\hline
\end{tabular}
\end{center}
\end{table}

\subsection{Impact of $r_{max}$}
As $r_{max}$ becomes shorter, the shape estimation becomes difficult. A problem appeared in the number of edge-vertex pairs because the number of observed edge-vertex pairs became small (Table \ref{comb_table_rmax50}). In particular, the number of observed edge-vertex pairs of the longest edge (c) is extremely small. This is because it becomes difficult to observe the whole edge of the longest edge by using a short $r_{max}$. Thus, we can no longer estimate the shape of $T$. Therefore, $r_{max}$ should be much longer than any of the edge lengths of $T$.

\begin{table}
\caption{Observed $\sharp\sensor_id(a,b)/E_{ind}[\sharp\sensor_id(a,b)]$ for a triangle $T$ with $r_{max}=50$}
\begin{center}\label{comb_table_rmax50}
\begin{tabular}{rlll}
\hline
Edge ID/vertex ID&A&B&C\\
a&1.143&1.241&0.7756\\
b&2.566&0.2786&0.8706\\
c&0.1996&0.3033&0\\
\hline
\end{tabular}
\end{center}
\end{table}

\subsection{Impact of sensing errors}
The proposed framework does not assume sensing errors, but sensing errors can exist in practice. Here, assume two types of sensing errors: one for $s_d$ and one for $l_d$. Sensing errors $\epsilon_s$ for $s_d$ are normally distributed with mean 0 and standard deviation $\sigma$. Due to the sensing error, $s_d$ becomes $\tan(\arctan(s_d)+\epsilon_s)$. The other type of sensing error divides $p_d$ into pieces. This type of error typically occurs when sensing reports are lost or slope changes are misjudged. Assume that errors of this type independently occur with probability $\epsilon_l$ at each sensing report during $p_d$. As a result, $l_d$ is divided into short $l_d$s.

\begin{figure}[tbh] 
\begin{center} 
\includegraphics[width=8cm,clip]{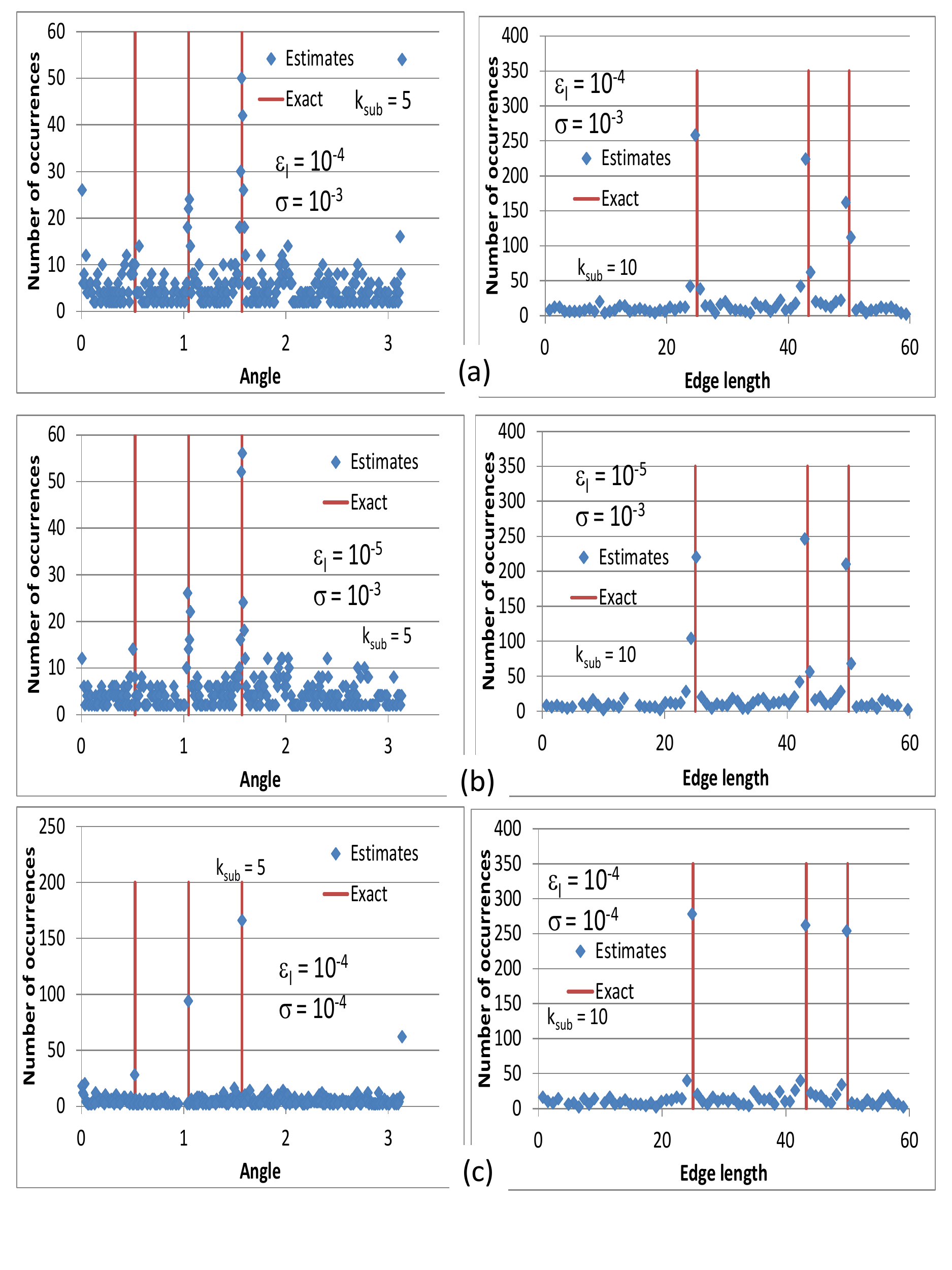} 
\caption{Impact of sensing errors: (a) $\epsilon_l=10^{-4}, \epsilon_s=10^{-3}$, (b) $\epsilon_l=10^{-5}, \epsilon_s=10^{-3}$, and (c) $\epsilon_l=10^{-4}, \epsilon_s=10^{-4}$.}
\label{noise}
\end{center} 
\end{figure}

Estimates under sensing errors are plotted in Fig. \ref{noise}. Sensing errors for $l_d$ ($s_d$) in Fig. \ref{noise}-(a) are more serious than those in Fig. \ref{noise}-(b) (Fig. \ref{noise}-(c)). For all cases except for the edge length estimates in (c), it is difficult to find three estimates. We cannot find clear sharp peaks of the number of candidate estimates, but there are many peaks. Even (c) and (a) contain a peak in angle estimates nearly $\pi$. This peak is caused by divided $l_d$, and each divided $l_d$ has a similar $s_d$. The proposed estimation method naturally judged there to be a vertex of angle nearly $\pi$ for consecutive edges providing similar $s_d$. Thus, this peak did not appear in (b) because of small $\epsilon_l$. As a whole, (c) looks better than (b). That is, accurate $s_d$ is needed to obtain good estimates for this example.

\subsection{Non-straight line driving route}
In this paper, a straight line driving route is assumed. Unfortunately, however, real driving routes are not straight. Here, a driving route has a $\pi/2$ right or left turn once in $\Omega$. The turn is randomly placed and goes right or left with probability 0.5. Such information is not available for estimation.

\begin{figure}[tb] 
\begin{center} 
\includegraphics[width=11cm,clip]{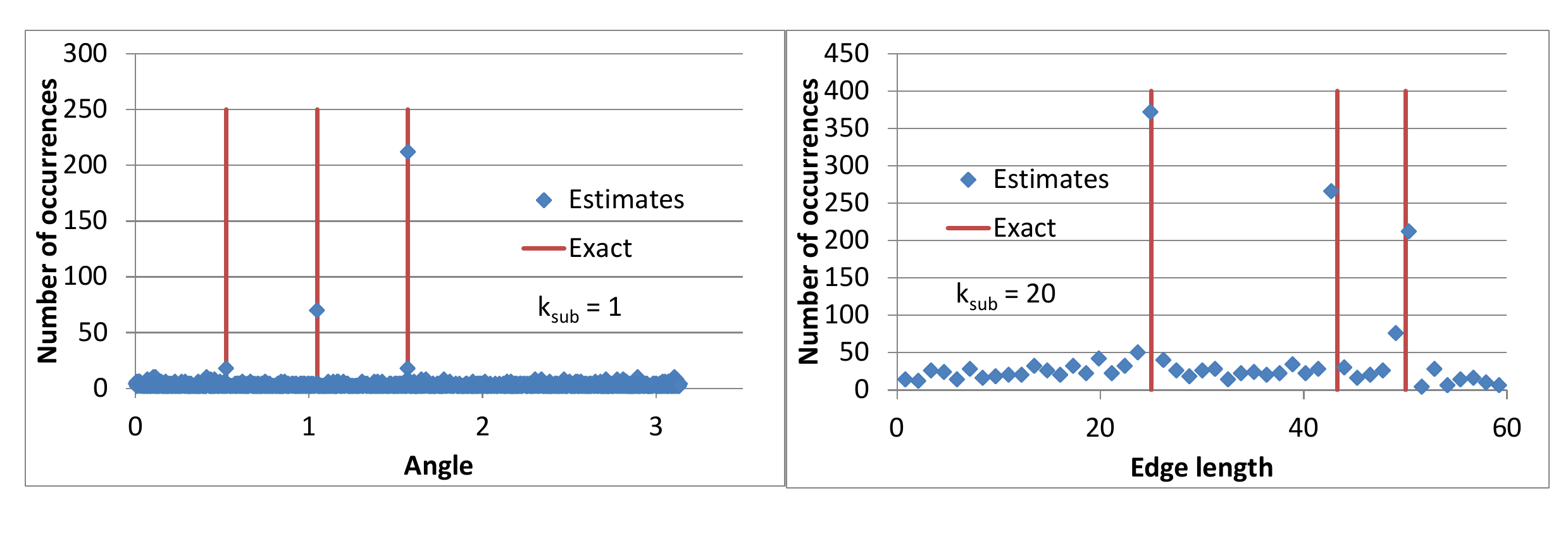} 
\caption{Estimation under non-straight driving route}
\label{non-straight}
\end{center} 
\end{figure}

The results are shown in Fig. \ref{non-straight}. We can find two peaks of angle estimates, but it becomes difficult to find an angle nearly $\pi/6$. If we pick out an angle nearly $\pi/6$, we should also pick out an angle nearly $\pi/2$. (There are two estimates nearly $\pi/2$ in this case.)

$\widehat{N_\lambda}$ is 1.468, 1.223, and 1.044 for the short, middle-length, and long edges, and is barely acceptable. On the other hand, $\widehat{N_{\gamma}}$ is overestimated particularly for $\widehat{\gamma}\approx\pi/2$ when we use two or four angle estimates. This is because the $\pi/2$ turn made the proposed method incorrectly estimate there to be many vertexes of angle $\pi/2$.

\subsection{Realistic target object}
\begin{figure}[tb] 
\begin{center} 
\includegraphics[width=11cm,clip]{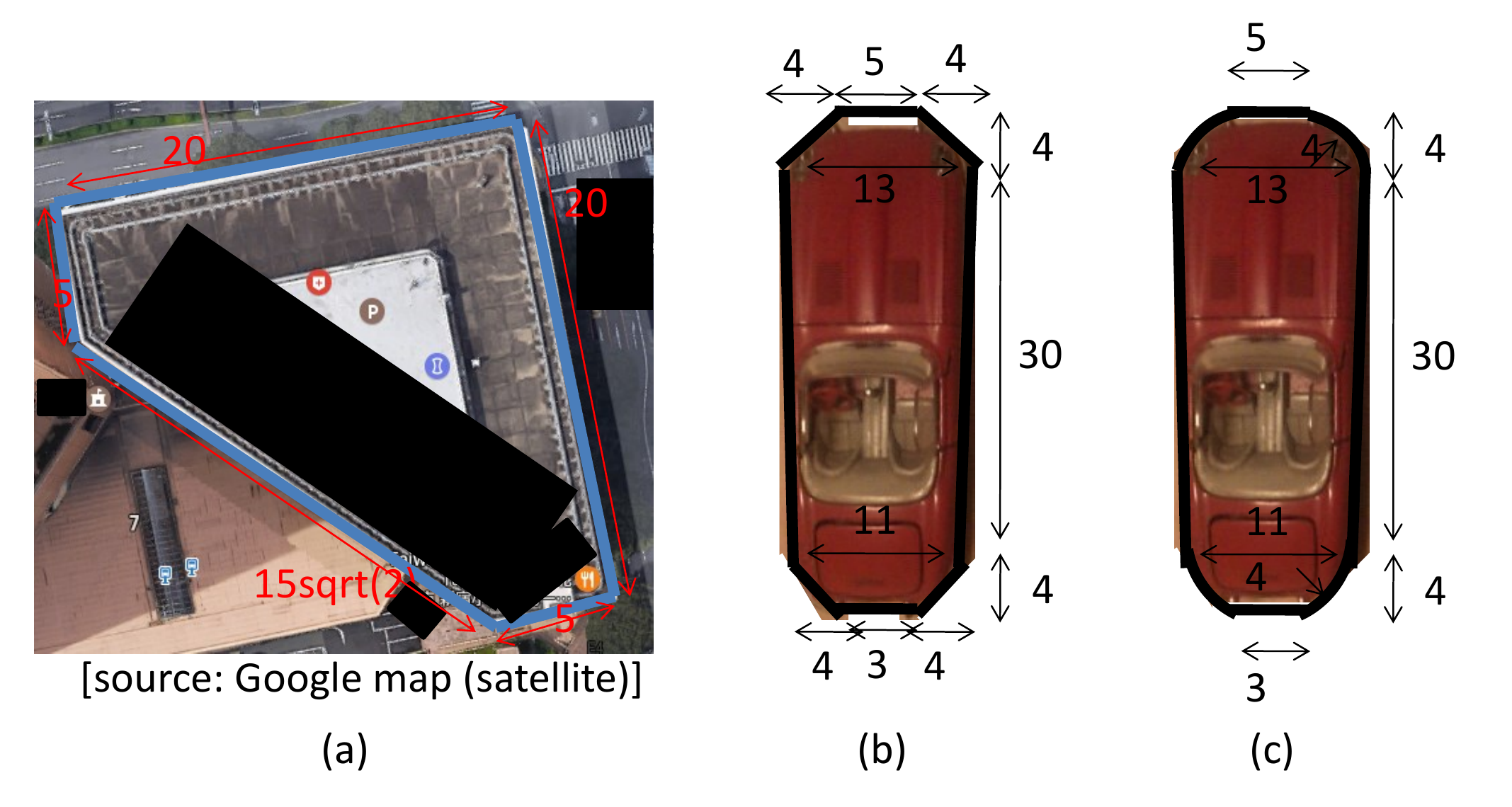} 
\caption{Realistic target objects}
\label{real_target}
\end{center} 
\end{figure}

Here, we estimate the shape of the building or cars shown in Fig. \ref{real_target}.

\subsubsection{Estimating building}
\begin{figure}[tb] 
\begin{center} 
\includegraphics[width=11cm,clip]{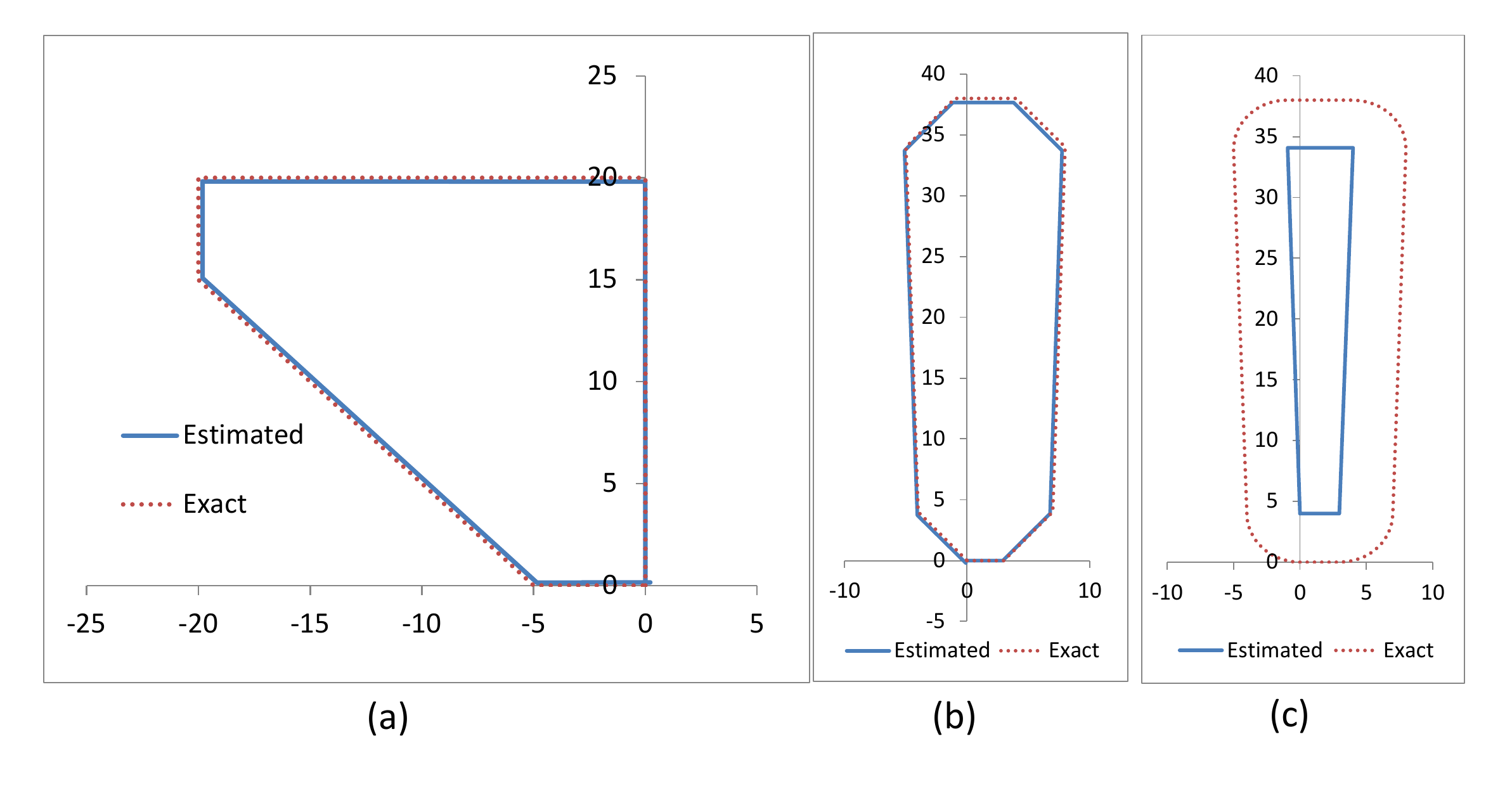} 
\caption{Estimated shape under default conditions}
\label{shape_result} 
\end{center} 
\end{figure}

Table \ref{result_building} shows that angle and edge length estimates were acceptable for the building in Fig. \ref{real_target} and that $\widehat{N_\lambda}$ was almost exact and that $\widehat{N_\gamma}$ was barely acceptable.
In addition, Table \ref{comb_building} suggests that edge (a) does not connect vertex (B) and that edge (d) does not connect vertex (A).
Because $\widehat{N_\lambda}\approx 2$ edge (a) and $\widehat{N_\lambda}\approx 1$ for other edges and $\widehat{N_\gamma} \approx 3$ for vertex (A) and $\widehat{N_\gamma}\approx 2$ for vertex (B), the shape of $T$ was obtained (Fig. \ref{shape_result}-(a)).

\begin{table}
\caption{Summary of estimated results for target object (a) under default conditions}
\begin{center}\label{result_building}
\begin{tabular}{rlll}
\hline
Edge ID&$\widehat{\lambda}$&Estimation error (\%)&$\widehat{N_\lambda}$\\
a&19.82&-0.9186&1.797\\
b&4.736&	-5.287&1.291\\
c&5.071&1.416&0.8639\\
d&21.16&-0.2660&0.7770\\
Vertex ID&$\widehat{\gamma}$&Estimation error (\%)&$\widehat{N_\gamma}$\\
A&1.571&-0.005921&3.420\\
B&2.356&-0.02124&1.654\\
\hline
\end{tabular}
\end{center}
\end{table}

\begin{table}
\caption{Edge-vertex combination for target object (a) under default conditions}
\begin{center}\label{comb_building}
\begin{tabular}{rll}
\hline
Edge ID/vertex ID&A&B\\
a&1.444&0.2291\\
b&0.7870&1.254\\
c&0.8915&1.454\\
d&0.4132&1.852\\
\hline
\end{tabular}
\end{center}
\end{table}

\subsubsection{Estimating polygon car}
We estimated car (b) in Fig. \ref{real_target}. The sum of the estimated number of edges was 7 for $n_s=1000$ and $\widehat{N_\lambda}\approx 1$ for $\widehat{\lambda}\approx 30$. Therefore, we failed to identify the shape of $T$.  This result seems to be because car (b) has more angles and edges (eight angles and edges) than other target objects in this paper. 

Thus, we estimated car (b) with $n_s=10,000$. The results are shown in Table \ref{result_car_b}. The estimation accuracy of $\widehat{\lambda}$ was fair, and that of $\widehat{\gamma}$ was good. In particular, the small difference in $\gamma$ was accurately estimated. $\widehat{N_\lambda}\approx 2$ for $\widehat{\lambda}\approx 30$ (edges (d) and (f)) and $\widehat{N_\lambda}\approx 1$ for $\widehat{\lambda}\approx 3$ (edge (c)) were desirable results. Although $\lambda =4$ and $\lambda =5$ were not clearly distinguished, $\widehat{N_\lambda}\approx 5$ for $\widehat{\lambda}\approx 4$ or 5 was an acceptable result. Because all the $\widehat{\gamma}$ were almost the same, we cannot identify the shape of $T$ (Remark \ref{shape_remark}). Although we cannot formally identify the shape, we can illustrate a car under some assumptions: it has an almost symmetrical shape and a head slightly wider than its tail. Two short edges connect vertexes (A), and the four consecutive edges of these edges are edges (a) or (b). One estimated shape is plotted in Fig. \ref{shape_result}-(b).  Its shape is not uniquely identified even under this assumption, but the estimated shape becomes almost the same as the actual shape. 

\begin{table}
\caption{Summary of estimated results for target object (b) with $n_s=10,000$}
\begin{center}\label{result_car_b}
\begin{tabular}{rlll}
\hline
Edge ID&$\widehat{\lambda}$&Estimation error (\%)&$\widehat{N_\lambda}$\\
a&5.593&	-1.1312&2.832\\
b&5.465&-3.390&1.066\\
c&2.911&-2.978&0.8249\\
d&29.86&-0.5221&0.9389\\
e&4.954&-0.9157&0.7653\\
f&29.99&-0.09659&0.8947\\
Vertex ID&$\widehat{\gamma}$&Estimation error (\%)&$\widehat{N_\gamma}$\\
A&2.356&-0.01143&3.701\\
B&2.389&0.2820&1.882\\
C&2.323&-0.3115&1.836\\	
\hline
\end{tabular}
\end{center}
\end{table}

\subsubsection{Estimating non-polygon car}
\begin{figure}[tb] 
\begin{center} 
\includegraphics[width=11cm,clip]{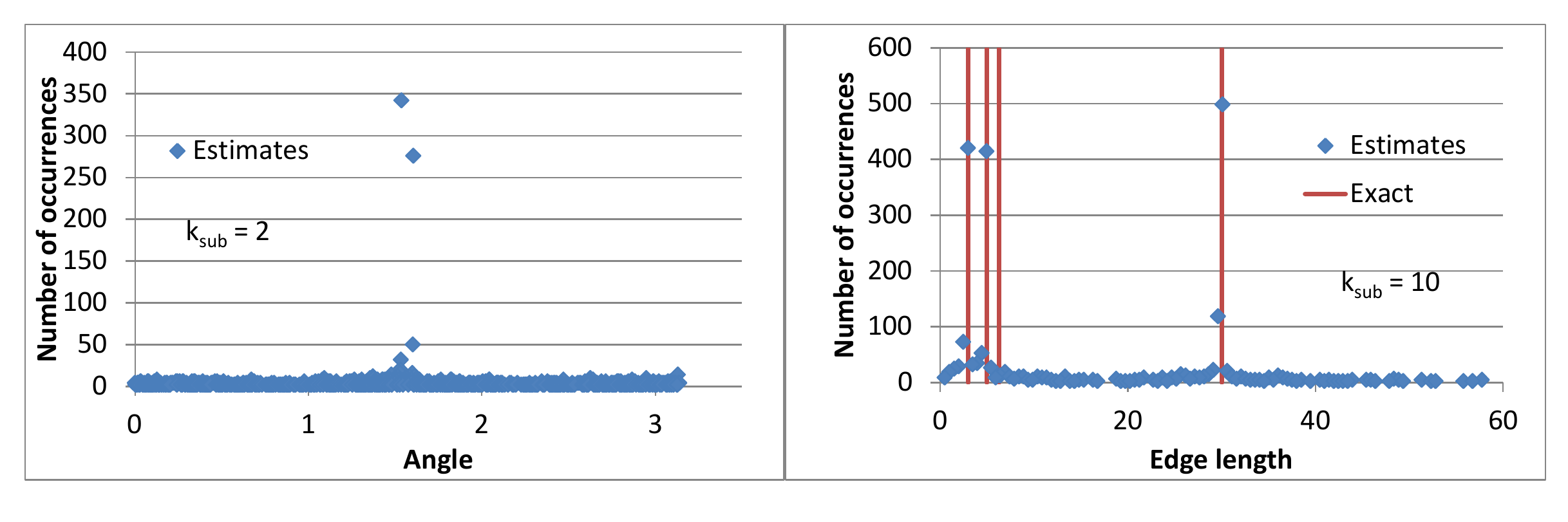} 
\caption{Estimated angles and lengths of target object (c) under default conditions}
\label{car}
\end{center} 
\end{figure}

Here, we show the estimated results for car (c) in Fig. \ref{real_target} with $n_s=1000$. Note that this target object is not a polygon. Figure \ref{car} plots the estimated angle and edge length. Table \ref{result_car} suggests that this target object is a quadrangle: two long edges (a), a single short edge (b), and a single short edge (c); two vertex (A) and two vertex (B). All the vertex angles are nearly $\pi/2$. These results are derived because the round corners of this target result in curves of $r(t)$. Because a curve in $r(t)$ can appear at a vertex of a polygon $T$ (Fig. \ref{curve}-(a)), the curve is not distinguishable from the curves at the round corners. Thus, the estimation method connected edges directly without round corners. Because the estimated angles of vertex (A) and vertex (B) are similar, an edge-vertex combination did not clearly estimate which edge connects vertex (A) and which edge connects vertex (B) (Remark \ref{shape_remark}). In Fig. \ref{shape_result}-(b), the estimated shape of this target object is plotted under the assumption that edge (b) connects two vertex (B). The estimated shape was almost identical to the shape with four corners of the original $T$ removed.

\begin{table}
\caption{Summary of estimated results for target object (b) under default conditions}
\begin{center}\label{result_car}
\begin{tabular}{rlll}
\hline
Edge ID&$\widehat{\lambda}$&Estimation error (\%)&$\widehat{N_\lambda}$\\
a&30.10&0.2854&1.522\\
b&2.971&-0.9795&1.029\\
c&4.944&-1.123&1.030\\
Vertex ID&$\widehat{\gamma}$&Estimation error (\%)&$\widehat{N_\gamma}$\\
A&1.538&--&	2.227\\
B&1.605&--&1.67\\
\hline
\end{tabular}
\end{center}
\end{table}

\section{Conclusion}
This paper proposed a theoretical framework for estimating target object shape by using distance sensors and speed meters mounted on vehicles. Here, the location and moving direction of these vehicles are unknown. Thus, the location privacy of vehicles is maintained. Several examples show that the proposed estimation framework is feasible. However, the proposed framework assumes some fairly strict conditions are met: the target object is a convex polygon, vehicles move on straight lines, and no sensing errors exist. Numerical examples suggest that the proposed framework may work even when these conditions are not satisfied, for example, vehicles move on a non-straight line or some sensing errors are imposed. For a non-polygon target object, the estimated shape becomes similar to the original target object shape without the non-polygon parts. Of course, we need additional effort to make the estimation method more robust under various conditions such as more serious erroneous sensing results and more frequent turns in a driving route. In addition, the proposed method requires many sensors for a complicated $T$. This is because we need to estimate the number of edges of a certain length or vertexes of a certain angle and because this number is large for a complicated $T$. Therefore, although the current proposed method does not use all the sensing data, a more efficient way of using sensing data should be developed.

A remaining large problem is the estimation for a non-convex target object.
For a non-convex target object, $r(t)$ becomes non-continuous and the analysis for $E[n_d(\widehat{\lambda})],E[n_d(\widehat{\gamma})]$ becomes complicated.
However, it seems possible to extend the proposed framework to cover a non-convex target object.
If so, we can also estimate the target object in the environment that there are many obstacles.
This is because we can estimate the total environment including the target object as a target object, although we cannot identify which is the original target object.

\appendix
\section{}
Here, we illustrate the reason the number of occurrences of other candidate estimates is less than the exact one under the assumption $\theta_{max}=\pi/2$.

Consider the edge length estimation. Here, we describe the case in which $-\theta_{max}\leq \zeta\leq\theta_{max}$. Then, the estimate $\widehat{\lambda}=l_d\sqrt{v^2-s_d^2}$ given by Eq. (\ref{lam1}) becomes exact. On the other hand, Eq. (\ref{lam2}) provides an incorrect estimate $l_d\sqrt{v^2+s_d^2}$. Thus, this incorrect estimate has no explicit relationship with the exact one and depend on $\phi$ through $s_d$.  Therefore, for various $\phi$, $l_d\sqrt{v^2+s_d^2}$ can be various.

Consider the vertex angle estimation.
For $\theta_{max}=\pi/2$, $\xi-\phi$ provided by Eq. (\ref{s_d1}) is $\arcsin(s_d/v), -\arcsin(s_d/v), \pi-\arcsin(s_d/v),\pi+\arcsin(s_d/v)$ and that by Eq. (\ref{s_d2}) is $\arctan(s_d/v), -\arctan(s_d/v), \pi-\arctan(s_d/v), \pi+\arctan(s_d/v)$.
In the remainder, $f(s_d),f'(s_d)=\arcsin(s_d/v), \arctan(s_d/v)$.
For simplicity, assume $s_d,s_d'\geq 0$ and $f(s_d)\geq f'(s_d')$.
Thus, the candidate estimates of $\pi-\gamma\in (0,\pi)$ are $f(s_d)-f'(s_d'), f(s_d)+f'(s_d'), \pi-f(s_d)+f'(s_d'), \pi-f(s_d)-f'(s_d')$.
This means that, if A is an candidate estimate of $\gamma$, $\pi$-A is also another candidate.

However, it is often the case that $\xi-\phi$ provided by Eq. (\ref{s_d1}) may not satisfy $-\theta_{max}\leq \zeta\leq\theta_{max}$ or  that provided by Eq. (\ref{s_d2}) may not satisfy $\theta_{max}<(>)\zeta$.
As a result, the candidate estimates of $\pi-\gamma$ are limited to, for example, $f(s_d)-f'(s_d'), \pi-f(s_d)-f'(s_d')$.
This means that even if A is an candidate estimate of $\gamma$, $\pi$-A is not and depends on $\phi$ through $s_d$.
 Therefore, for various $\phi$, $\pi$-A can be various.

\end{document}